%% file: main.tex
\documentclass[sigconf, nonacm]{acmart}

\newcommand{\SystemName}{Helios}
\title{\SystemName: An Efficient Out-of-core GNN Training System on Terabyte-scale Graphs with In-memory Performance}
\usepackage{tikz}
\newcommand*\circled[1]{\tikz[baseline=(char.base)]{
            \node[shape=circle,fill,color=black,text=white,inner sep=0.05pt](char){#1};}}

\usepackage{enumitem}
\usepackage{subfigure}

\begin{document}

\author{Jie Sun, Mo Sun, Zheng Zhang, Jun Xie, Zuocheng Shi, Zihan Yang, Jie Zhang, Fei Wu, Zeke Wang}
\affiliation{%
  \institution{Zhejiang University, China}
    \institution{Shanghai Institute for Advanced Study of Zhejiang University, China}
}
\email{{jiesun,sunmo,zhzh2001,junxiecs,shizuocheng,zihanyang,carlzhang4,wufei,wangzeke}@zju.edu.cn}

\input{context/abstract.tex}

\maketitle

\vspace{-1ex}
\input{context/introduction.tex}

\vspace{-1ex}
\input{context/motivation.tex}
\vspace{-1ex}
\input{context/design.tex}

\vspace{-1ex}
\input{context/evaluation.tex}
\vspace{-1ex}
\input{context/relatedwork.tex}

\vspace{-1ex}
\input{context/conclusion.tex}

\bibliographystyle{ACM-Reference-Format}
\bibliography{references}

\end{document}

%% file: context/abstract.tex
\begin{abstract}
Training graph neural networks (GNNs) on large-scale graph data holds immense promise for numerous real-world applications but remains a great challenge.
Several disk-based GNN systems have been built to train large-scale graphs in a single machine. However, they often fall short in terms of performance, especially when training on terabyte-scale graphs. This is because existing disk-based systems either overly focus on minimizing the number of SSD accesses or do not fully overlap SSD accesses with GNN training, thus resulting in substantial unnecessary overhead on the CPU side and then low GPU utilization. 
To this end, we propose \SystemName{}, a system that can train GNN on terabyte graphs in a single machine while achieving throughput comparable with in-memory systems. To achieve this, we first present a GPU-initiated asynchronous disk IO stack, allowing the GPU to directly access graph data on SSD. This design only requires about 30\% GPU cores to reach the almost maximal disk IO throughput and wastes no GPU cores between IO submission and IO completion such that the majority of GPU cores are left for other GNN kernels. Second, we design a GPU-managed heterogeneous cache that extends the cache hierarchy to heterogeneous CPU and GPU memory and thus enhances cache lookup throughput significantly by GPU parallelism. Finally, we build a deep GNN-aware pipeline that seamlessly integrates the computation and communication phases of the entire GNN training process, maximizing the utility of GPU computation cycles. 
Experimental results demonstrate that \SystemName{} can match the training throughput of in-memory GNN systems, even for terabyte-scale graphs. Remarkably, \SystemName{} surpasses the state-of-the-art GPU-managed baselines by up to 6.43$\times$ and exceeds CPU-managed baselines by over 182$\times$ on all terabyte-scale graphs. 

\end{abstract}

%% file: context/introduction.tex
\section{Introduction}
Graph neural network (GNN)~\cite{kipf2016semi, hamilton2017inductive,  velivckovic2017graph, chiang2019cluster, zeng2019graphsaint, chen2018fastgcn} is a category of deep learning models designed to train on both structural and attribute data of graphs to generate low-dimensional embeddings. These embeddings are instrumental in executing machine learning tasks such as node classification and link prediction. GNNs have proven efficacious in a myriad of practical domains, such as recommendation systems in e-commerce platforms, fraud detection, risk control in financial management, and molecular property prediction in drug development~\cite{ying2018graph, liu2021pick, yuapplication, su2022gnn, gilmer2017neural}. 
% Many GNN systems~\cite{jia2020improving, wang2019deep,fey2019fast, zhu2019aligraph, lin2020pagraph, yang2022gnnlab, torchquiver, ma2019neugraph, zhang2020agl, wang2021gnnadvisor, gandhi2021p3, thorpe2021dorylus, zheng2022distributed, sun2023legion, zhang2023boosting} are proposed to ease the development and accelerate the training of GNN models.

Many existing large-scale GNN systems~\cite{sun2023legion, zhang2023ducati, park2022ginex, waleffe2023mariusgnn, park2023accelerating, liu2021bgl, zheng2020distdgl, zheng2021distributed, gandhi2021p3, gong2023gsampler} adopt sampling-based GNN models~\cite{hamilton2017inductive, zeng2019graphsaint, chen2018fastgcn} to extend GNN training to very large graphs. Among them, several recent works~\cite{park2022ginex, waleffe2023mariusgnn, park2023accelerating} have explored disk-based GNN training in a single machine for much less monetary cost~\cite{waleffe2023mariusgnn} compared to the distributed in-memory solutions~\cite{liu2021bgl, zheng2020distdgl, zheng2021distributed, gandhi2021p3}. Training GNNs on terabyte-scale graphs in a single server remains a great challenge, because the size of graph data can easily exceed the upper limit of CPU memory capacity in a modern server. For example, in Alibaba's Taobao recommendation system, the user behavior graph contains more than one billion vertices and tens of billions of edges~\cite{zhu2019aligraph}, which need several terabytes of storage space. Despite the promising future of GNN applications, we still identify that existing disk-based GNN systems face severe performance issues especially when training on terabyte-scale graphs. To illustrate the issues, we categorize these systems into two types according to the mini-batch preparation mechanism:

\begin{figure*}[t]
    \subfigure[Execution time breakdown of existing CPU-managed systems when training on the PA dataset~\cite{hu2020open}. ]{
        \label{fig_motivation_1}
        \includegraphics[width=0.32\linewidth]{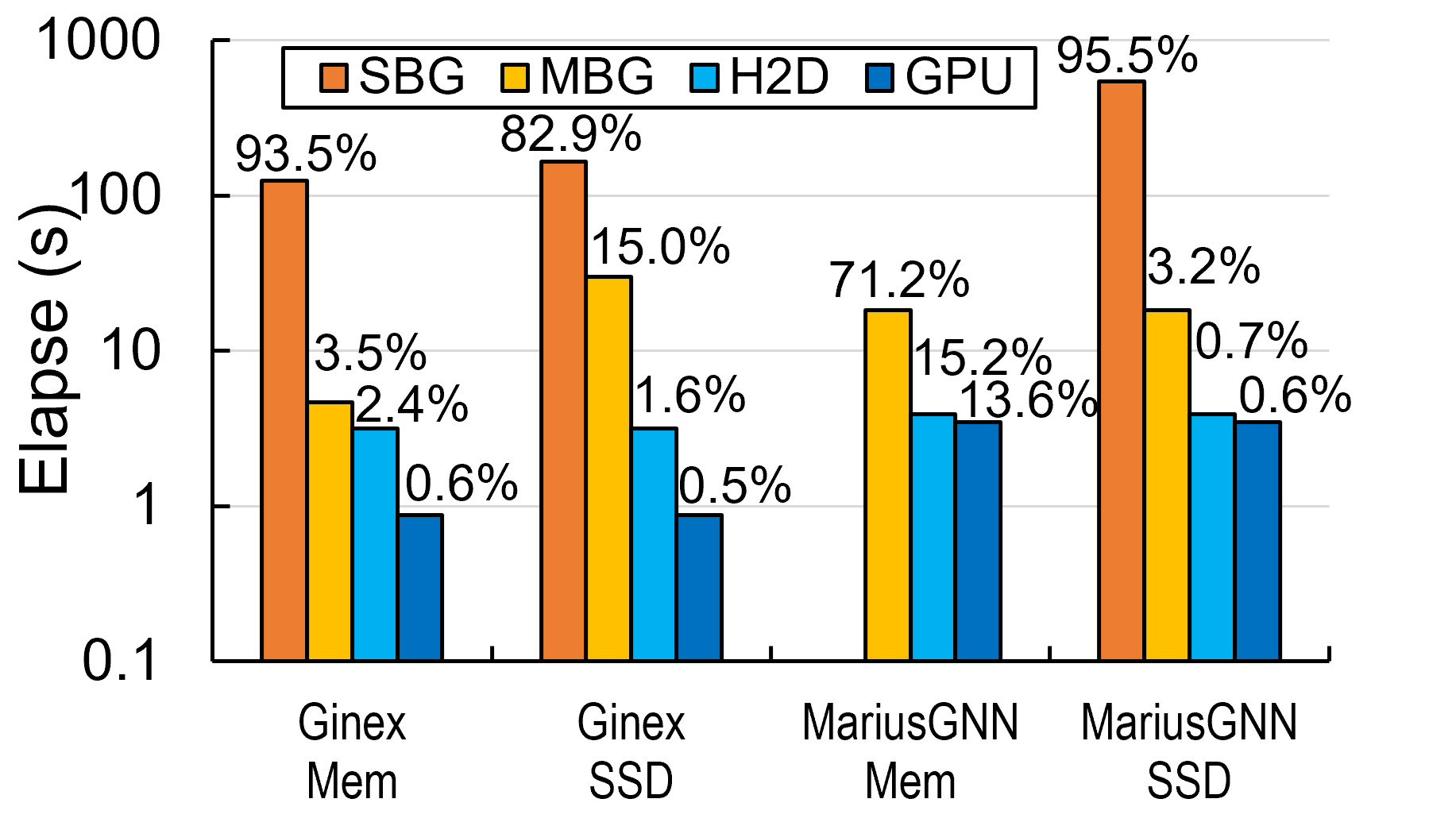}
    }
    \hfill    
    \subfigure[GPU PCIe utilization under the CPU-managed cache.]{
        \label{fig_motivation_2}
        \includegraphics[width=0.32\linewidth]{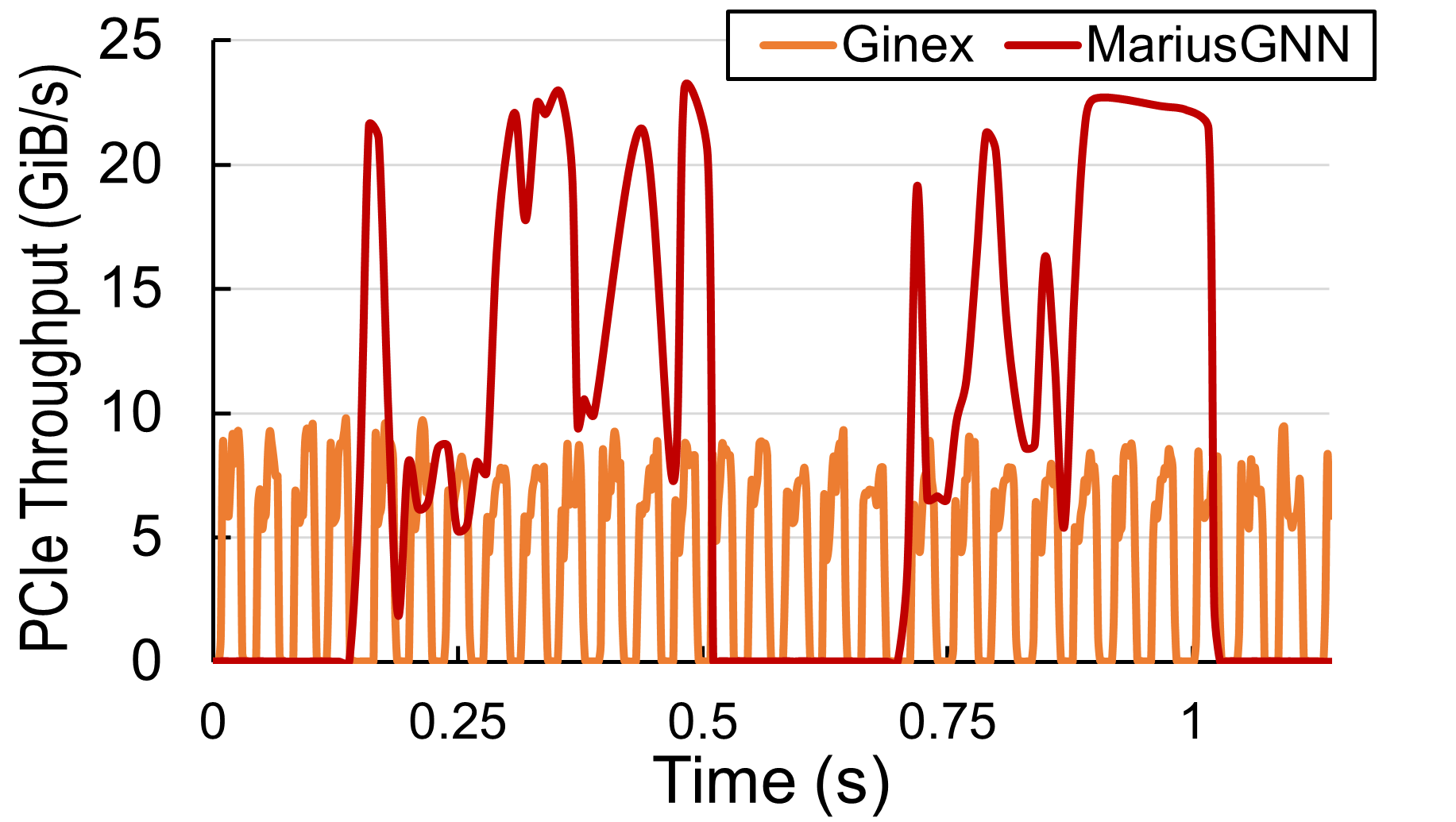}
    }
    \hfill    
    \subfigure[IO throughput achieved by GIDS feature extraction. Feature dimension is 1024. ]{
        \label{fig_motivation_io}
        \includegraphics[width=0.32\linewidth]{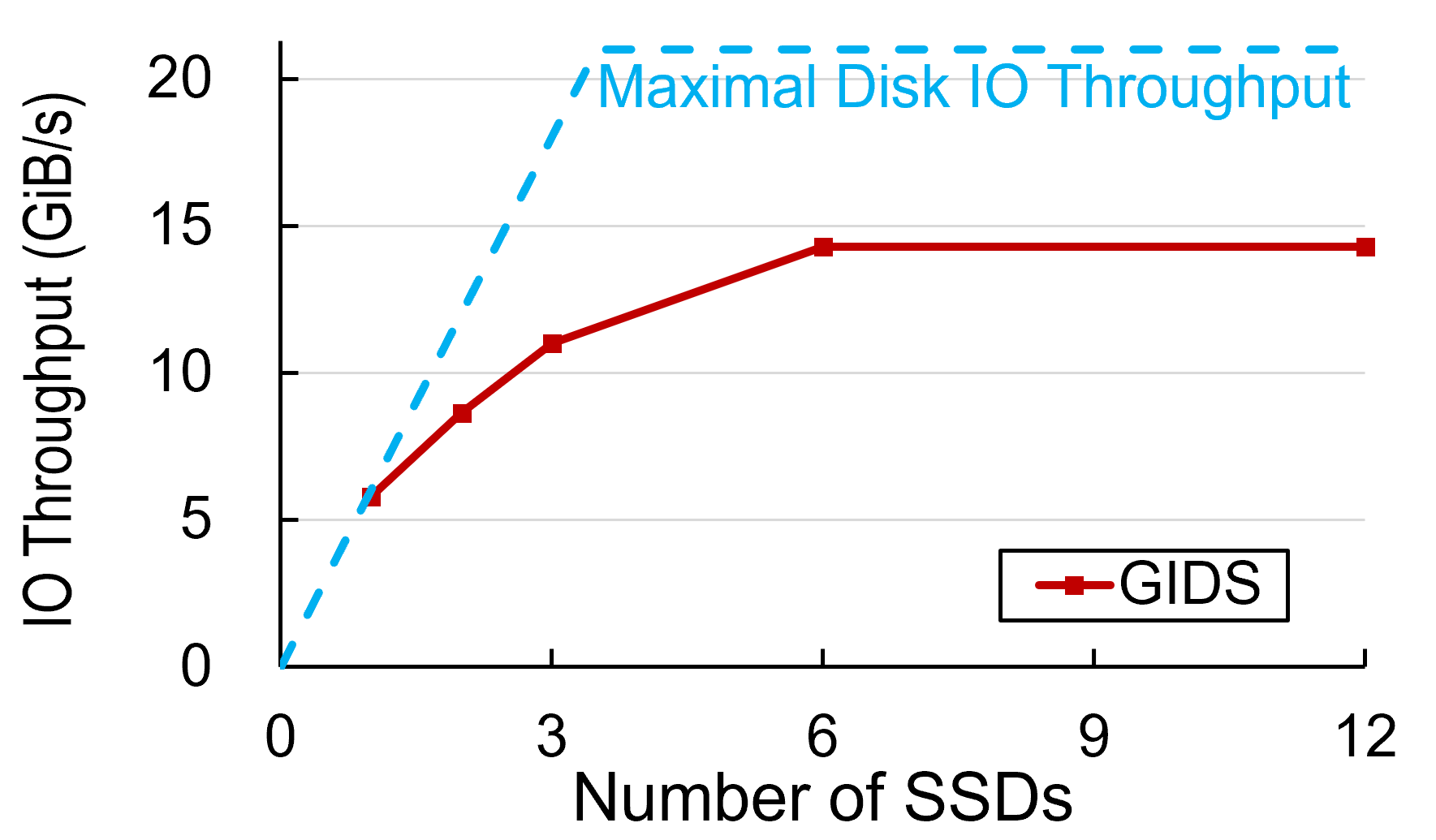}
    }
\vspace{-3ex}   
\caption{Issues of existing CPU- and GPU-managed Systems motivate the design of \SystemName. SBG, MBG, H2D, and GPU represent the super-batch generation, the mini-batch generation, copying mini-batch to GPU, and model training on GPU, respectively.} 
   \vspace{-3ex}
    \label{fig_motivation} 
\end{figure*} 

\begin{itemize}[topsep=5pt, leftmargin=*]
\item {\bf CPU-managed Systems. } Ginex~\cite{park2022ginex} and MariusGNN~\cite{waleffe2023mariusgnn} leverage the CPU to execute a super-batch generation to fill up CPU memory with pre-sampled data or graph partitions and thereby minimize subsequent SSD accesses during the corresponding mini-batch generations, e.g., 1500 mini-batches per super-batch. In each mini-batch generation, these systems use the CPU to prepare sampled vertices as well as their features and copy the mini-batch to GPU for training. However, we identify that these two systems are severely bottlenecked by CPU processing and have down to 0.6\% GPU utilization. The reasons are twofold: 1) excessive CPU workload on super-batch and mini-batch preparation that dominates over 70\% training time, leading to serial execution of CPU and GPU and low GPU utilization (See Figure~\ref{fig_motivation_1}); and 2) serial execution of CPU-managed cache/buffer lookup and CPU-GPU data transfer, resulting in low GPU PCIe utilization (See Figure~\ref{fig_motivation_2}). 

\item {\bf GPU-managed System. }
GIDS~\cite{park2023accelerating} uses GPU to prepare mini-batch data by graph sampling and feature extraction. And GIDS utilizes BaM~\cite{qureshi2023gpu} system to manage disk IO. This solution can avoid CPU bottlenecks by leveraging GPU's parallelism. However, GIDS has low overall training throughput due to three severe issues: 1) failing to maximize the disk IO throughput (Down to 60\% as shown in Figure~\ref{fig_motivation_io}) due to insufficient parallelism, even though using all GPU cores; 2) exhausting all GPU cores under graph sampling, feature extraction, and model training, leading to almost serial execution of all these stages; 3) having a limited GPU cache space deficient for terabyte graphs, resulting in over 73\% proportion of data still fetched from disks. 
\end{itemize}

In essence, the existing disk-based GNN systems do not fully leverage the performance potentials of SSDs, even at the cost of consuming all the GPU resources. It is widely accepted that a computing task that involves massive data movement from disk could be significantly slower than a computing task that directly accesses CPU memory. Therefore, these systems either prompt pronounced efforts to minimize SSD accesses, which results in substantial data preparation overhead on the CPU side, or do not fully overlap direct SSD accesses with GNN training, which causes low GPU utilization. In this paper, we ask: 
\begin{center}
{\em Can we train a GNN model on a terabyte-scale graph stored in disks while still achieving in-memory throughput?}
\end{center}

To this end, we propose \SystemName{}, an efficient out-of-core GNN training system that can train GNN on terabyte-scale graphs with in-memory performance. The key idea is to minimize the GPU resource overhead for direct SSD accesses, and thus have sufficient GPU resources for heterogeneous cache-enabled in-memory GNN training that fully pipelines with direct SSD accesses and minimizes the number of costly SSD accesses. 
%fully overlap GPU GNN training and data access from SSDs, while leveraging efficient SSD control and cache design using GPU parallelism. 
% To achieve this, we employ GPU-initiated operators coupled with meticulously allocated GPU parallelism and schedule them with deep pipeline. 
Specifically, \SystemName{} consists of three key innovations:

\begin{itemize}[topsep=5pt, leftmargin=*]
\item {\bf GPU-initiated Asynchronous Disk IO Stack. }We propose a GPU-initiated asynchronous disk IO stack that consists of decoupled thread-level parallel IO command submission and asynchronous IO completion handling. In our design, the IO stack only requires \textasciitilde 30\% GPU cores to submit sufficient IO requests to reach the almost maximal IO throughput of multiple SSDs. Moreover, the IO completion handling is decoupled from IO submission, thus wasting no GPU cores between these two stages. Such a design leaves the majority of GPU cores for other GNN kernels. To the best of our knowledge, \SystemName{} is the first system that utilizes GPU to initiate decoupled asynchronous IO to directly access disks on demand without CPU involvement. 

\item {\bf GPU-managed Heterogeneous Cache. }We propose a GPU-managed heterogeneous cache that extends the cache hierarchy to heterogeneous CPU and GPU memory, minimizing low-throughput disk IO accesses. Meanwhile, we leverage GPU's massive parallelism to initiate cache lookup, boosting cache lookup throughput. The cache design further augments the GPU PCIe throughput, thus enabling \SystemName{} to achieve high throughput comparable with the in-memory system, even when the graph scale reaches terabytes.

\item {\bf Deep GNN-aware Pipeline. }We propose a deep GNN-aware pipeline that can fully utilize GPU computations cycles and thus maximize the overall throughput. The pipeline scheduler in \SystemName{} intelligently decomposes the comprehensive GNN training procedure into a sequence of GPU-initiated operators. It allocates suitable GPU resources to each operator and executes them according to the meticulously crafted execution plan. 
\end{itemize}

We implement \SystemName{} on the top of the SOTA in-memory system Legion~\cite{sun2023legion}. We evaluate \SystemName{} on popular GNN models~\cite{hamilton2017inductive, velivckovic2017graph} and real-world graphs~\cite{hu2020open, khatua2023igb, BoVWFI, BRSLLP, BCSU3, BMSB, angles2020ldbc} whose sizes are up to 23TB. Experiments show that \SystemName{} can achieve 90\%-99\% training throughput compared to that of in-memory GNN systems on small datasets (See Figure~\ref{fig_in_memory}) while \SystemName{} maintains the same GPU PCIe throughput for terabyte-scale graphs (See Figures~\ref{fig_cache_ssdnum},~\ref{fig_cache_featdim}). Remarkably, \SystemName{} outperforms the state-of-the-art GPU-managed baselines by up to a factor of 6.43$\times$ and exceeds CPU-managed counterparts by over 182$\times$ on terabyte-scale graphs (See Figure~\ref{fig_endtoend}).

%% file: context/motivation.tex
\section{Motivation}

\label{motivation_sec}
A series of systems~\cite{park2022ginex, waleffe2023mariusgnn, park2023accelerating} have been proposed to enable disk-based GNN training on large-scale graphs. 
However, these existing disk-based GNN systems do not fully leverage the performance potentials of SSDs. It is widely accepted that a computing task that involves massive data movement from disk could be significantly slower than a computing task that directly accesses CPU memory. Therefore, these systems either prompt pronounced efforts to minimize SSD access, which results in substantial data preparation overhead on CPU, or do not fully overlap SSD access with GNN training, which causes low GPU utilization.
We will discuss the concrete performance issues in this Section. According to the mini-batch preparation mechanism, we classify these systems into two types: CPU-managed systems~\cite{park2022ginex, waleffe2023mariusgnn} and GPU-managed systems~\cite{park2023accelerating}. 

\subsection{Issues of CPU-managed Systems}
Ginex~\cite{park2022ginex} and MariusGNN~\cite{waleffe2023mariusgnn} use CPU to prepare mini-batch data and use GPU for model training. We find that these CPU-managed designs are severely bottlenecked by CPU processing, leading to under-utilization of GPU. The concrete reasons are twofold:

\noindent{\bf I$_1$: Low GPU Utilization Due to the Serial Execution of CPU and GPU. }
Ginex~\cite{park2022ginex} and MariusGNN~\cite{waleffe2023mariusgnn} place much computation-intensive memory-hungry workload on the CPU, thus leading to CPU bottlenecks and GPU under-utilization. The end-to-end execution in both two systems consists of four stages: 1) super-batch generation; 2) mini-batch generation; 3) copying mini-batch from CPU to GPU; and 4) training model on GPU. 

Specifically, Ginex~\cite{park2022ginex} adopts an inspector-executor model in each GNN training epoch. During the inspector stage, Ginex uses the CPU to process a super-batch including sampled vertices of all corresponding mini-batches, and execute feature cache management including loading features from disk to CPU memory. This stage is designed to mitigate SSD access by maximizing cache hit rate. During the executor stage, Ginex runs a main loop: 1) using the CPU to extract feature data from the disk and CPU cache, 2) copying data to GPU, and 3) training models on GPU. Similarly, MariusGNN~\cite{waleffe2023mariusgnn} generates super-batches to avoid SSD access during the main loop of mini-batch iteration. During the super-batch generation, MariusGNN uses the CPU to load graph partitions from disk to CPU memory. After loading partitions, MariusGNN runs a main loop including 1) generating mini-batch data by CPU in each in-memory partition and 2) sending data to GPU for training. 

% Ginex and MariusGNN finish one training epoch when all super-batch are traversed. 

We break down the end-to-end execution of Ginex and MariusGNN when training the PA~\cite{hu2020open} dataset, as shown in Figure~\ref{fig_motivation_1}. Specifically, Ginex-Mem and MariusGNN-Mem represent storing all topology and feature data in CPU memory while Ginex-SSD and MariusGNN-SSD represent storing all feature data in SSD and storing all topology data in CPU memory. We can see the CPU-managed super-batch and mini-batch generation processes occupy the 70-98\% proportion of total execution time in these systems, leading to very low utilization of GPU due to their serial execution on CPU and GPU.

\noindent{\bf I$_2$: Low GPU PCIe Utilization. } 
Both Ginex~\cite{park2022ginex} and MariusGNN~\cite{waleffe2023mariusgnn} leverage the CPU memory as a cache/buffer. They rely on CPUs to collect sparse features from a large feature table of cache/buffer to generate mini-batch data and use DMA to send them to GPU. Such a strategy leads to the serial execution of CPU-managed cache/buffer lookup and CPU-GPU data transfer, thus leading to low utilization of GPU PCIe. (Similar to the issues studied by in-memory GNN systems~\cite{min2021large}.) Figure~\ref{fig_motivation_2} illustrates the GPU PCIe throughput\footnote{PCIe throughput is collected by reading the performance counters of the PCIe switch port which the GPU attached to.} during the main loop of Ginex and MariusGNN when training on the PA dataset. 
To examine the impact of CPU-managed cache, we allocate sufficient cache/buffer size to ensure that all features can be accommodated in CPU memory. Observations from Figure~\ref{fig_motivation_2} reveal that the GPU PCIe is frequently under-utilized. It becomes evident that if Ginex and MariusGNN are required to extract features from SSDs, the utilization of GPU PCIe will be even lower.

\subsection{Issues of GPU-managed Systems}
GIDS~\cite{park2023accelerating} relies on GPU to initiate graph sampling, feature extraction, and model training without CPU involvement, avoiding CPU bottlenecks. To enable GPU-initiated SSD access, GIDS adopts the BaM~\cite{qureshi2023gpu} system's disk IO stack and uses the BaM software cache for vertex feature lookup. However, we still identify that GIDS has low training throughput due to three severe issues.

\noindent{\bf I$_3$: Failing to Maximize Disk IO Throughput.} Figure~\ref{fig_motivation_io} shows that GIDS achieves low throughput (down to 60\%) compared to the maximal disk IO throughput. We identify that GIDS fails to maximize disk IO throughput because of insufficient parallelism exploited by its design. 

Figure~\ref{iostack_bam} shows how GIDS extracts vertices' features. GIDS utilizes one GPU IO kernel to handle the whole feature extraction procedure, in which each warp extracts one vertex's feature.
The detailed process of the IO kernel consists of four steps. First, GIDS assigns each warp to look up BaM~\cite{qureshi2023gpu}'s page cache for one feature reading request (\circled{1}). Second, if getting a cache miss, a leader thread in the warp converts one request to one NVMe command\footnote{NVMe is a protocol used by current high-performance SSDs.} and submits the command to a submission queue (SQs) of SSDs (\circled{2}). After receiving the commands, SSDs will prepare the feature data and send them to the temporary IO buffer in GPU memory. Third, the leader thread pools completion queues (CQs) of SSDs until it gets a completion entry of a command (\circled{3}). After getting the completion entry, all threads in the warp move features from the temporary IO buffer to the output feature buffer (\circled{4}). 

According to the nature of SSDs, the key to maximizing disk IO throughput is to send enough concurrent NVMe commands in parallel. However, GIDS fails to do so due to two concrete reasons. First, most of the threads in a warp are not utilized. During the step (\circled{1}-\circled{3}), only one leader thread executes command submission and completion pooling, while all other threads are in a waiting state. This leads to most threads occupying GPU cores but doing nothing. Second, due to BaM~\cite{qureshi2023gpu}'s synchronous IO stack design, each warp can not execute other operations between the IO submission (\circled{2}) and the IO completion polling (\circled{3}). Since the disk IO has a very long latency, the parallelism of the GPU is severely limited. As such, though GIDS launches all GPU cores for the feature extraction kernel (256K thread blocks, 32 threads in each block, 100\% GPU core utilization\footnote{Measured with the SM Active metric via NVIDIA NSight Systems~\cite{nsight}.}), it can still hardly reach the maximal disk IO throughput (See Figure~\ref{fig_motivation_io}). 

% Unlike computing tasks, the execution time of the storage IO stack is bounded by SSD IO. Therefore, for completion pooling, GPU threads are in a waiting state before data transfer between GPU and SSDs is finished, wasting precious computing resources.

\noindent{\bf I$_4$: Almost Serial Execution of Graph Sampling, Feature Extraction, and Model Training. } 
According to the discussion above, GIDS relies on all GPU cores for initiating disk IO commands. Additionally, GIDS also relies on all GPU cores to execute graph sampling and model training. For example, GIDS launches kernels with 49K, 321K, and up to 24K thread blocks for graph sampling, feature extraction, and model training on the IG~\cite{khatua2023igb} dataset, respectively. As a result, each stage exhausts GPU cores, leading to the almost serial execution of all three stages. Thus, GIDS has low training throughput.

\noindent{\bf I$_5$: Large Cache Miss Due to Limited GPU Cache Space. }GIDS~\cite{park2023accelerating} reuses the GPU software cache in BaM~\cite{qureshi2023gpu} to reduce data transfer from the secondary storage. However, this cache space is strictly limited by rare GPU memory size, leading to a large cache miss rate when training on terabyte-scale graphs. For example, we evaluate GIDS in an 80GB A100 GPU, the maximal cache ratio is 3.6\% when training on the IG dataset (the smallest terabyte-scale graph in Table~\ref{dataset}), leading to about 73\% cache miss rate. 

%% file: context/design.tex
\section{Design of \SystemName{}}

\begin{figure}[t]
    \vspace{-2ex}
    \begin{center}
        \subfigure[\label{iostack_bam} In GIDS]{
            \includegraphics[width=0.477\linewidth]{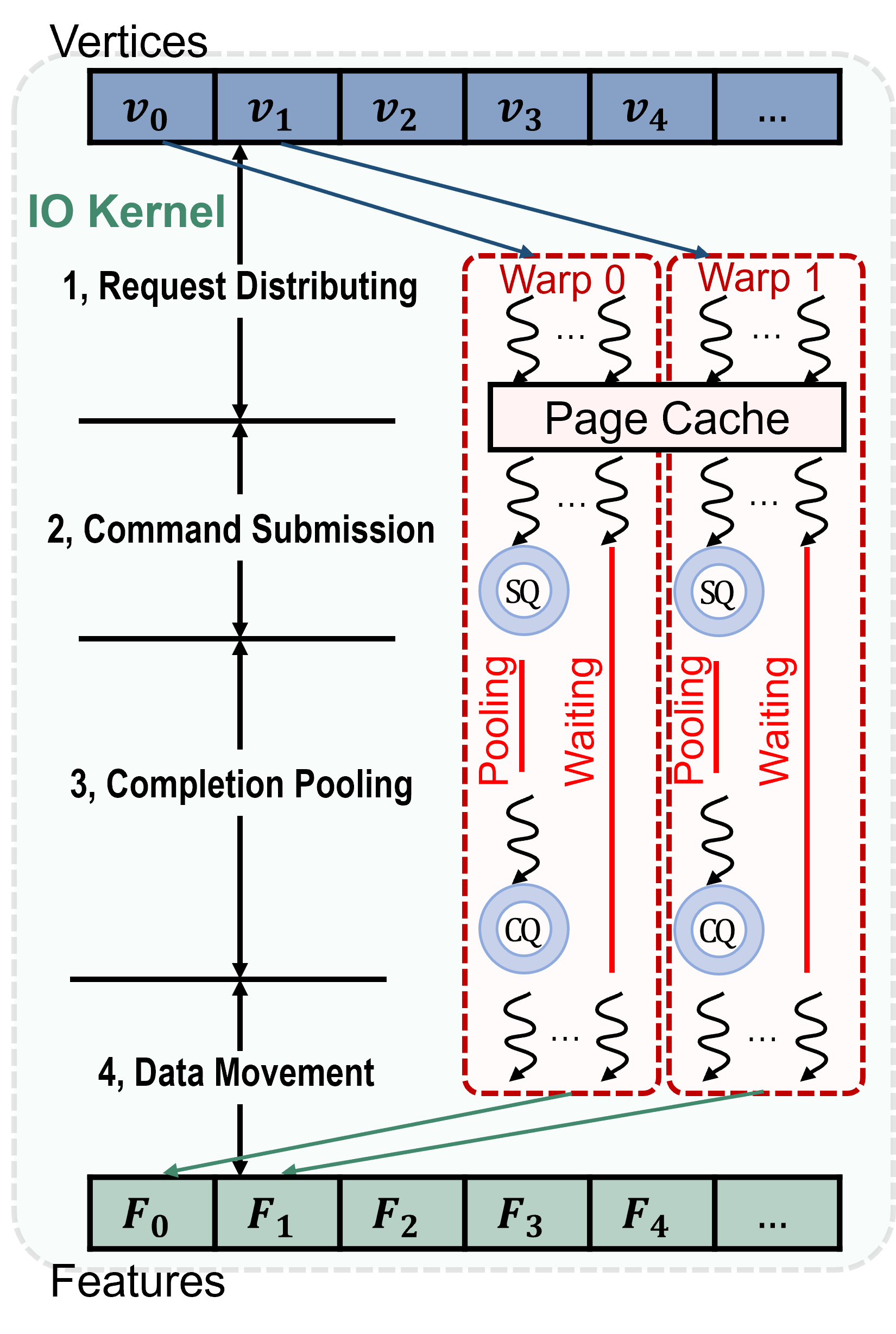}
        }
        \subfigure[\label{iostack_ours} In \SystemName{}]{
            \includegraphics[width=0.473\linewidth]{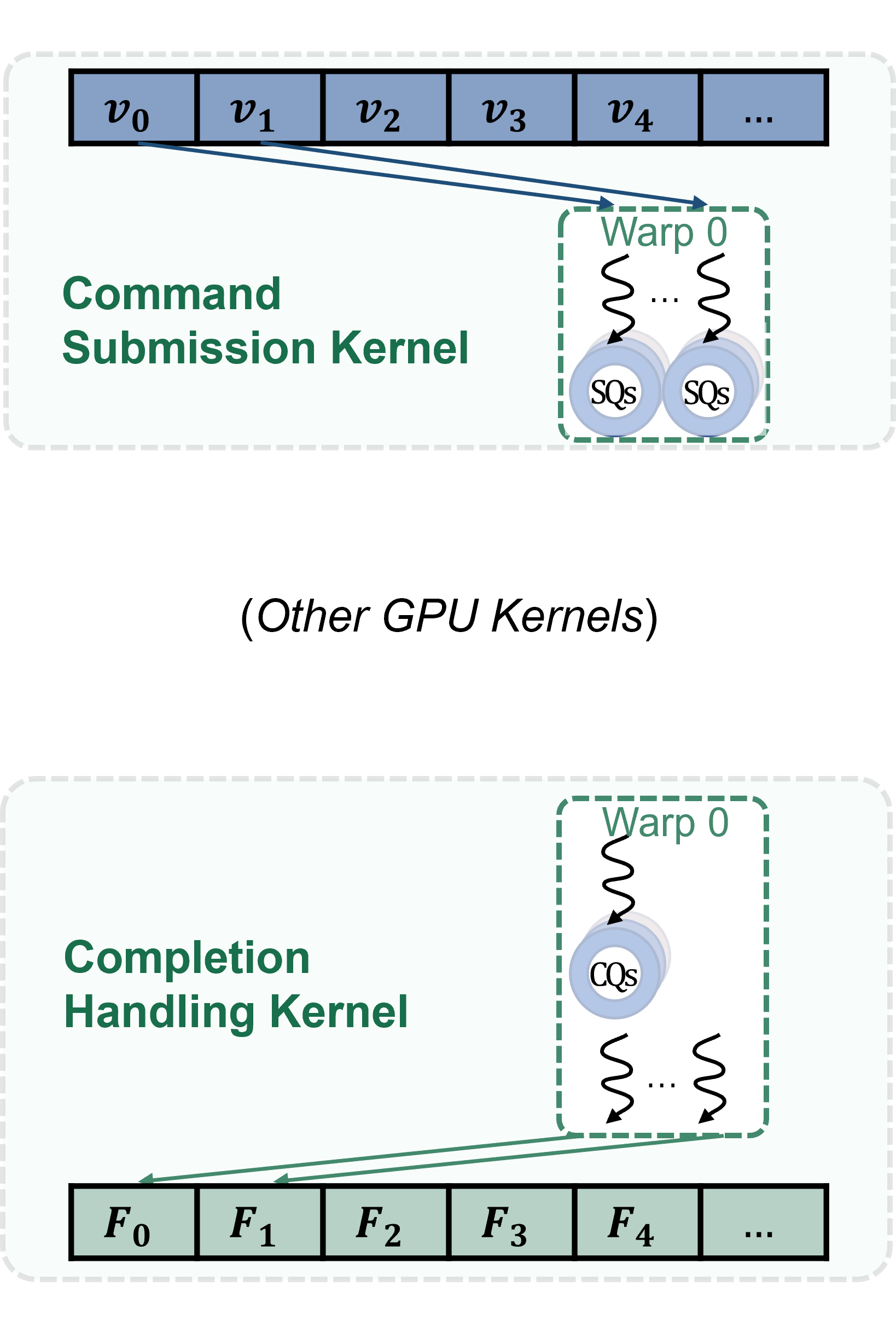}
        }
    \end{center}
    \vspace{-2ex}
    \caption{\label{iostack} Comparison of GIDS and \SystemName: IO process of each GPU warp in GPU-initiated systems. GIDS occupies all the GPU cores during IO access, while \SystemName{} only needs a small amount of GPU cores to launch and monitor the completion of IO completion, leaving the majority of GPU cores for computation. $v_i$ represents vertex $i$ and $F_i$ represents feature of vertex $i$.}
    \vspace{-4ex}
\end{figure}

\begin{figure}[t]
\vspace{-0ex}
	\centering
	{\includegraphics[width=\linewidth]{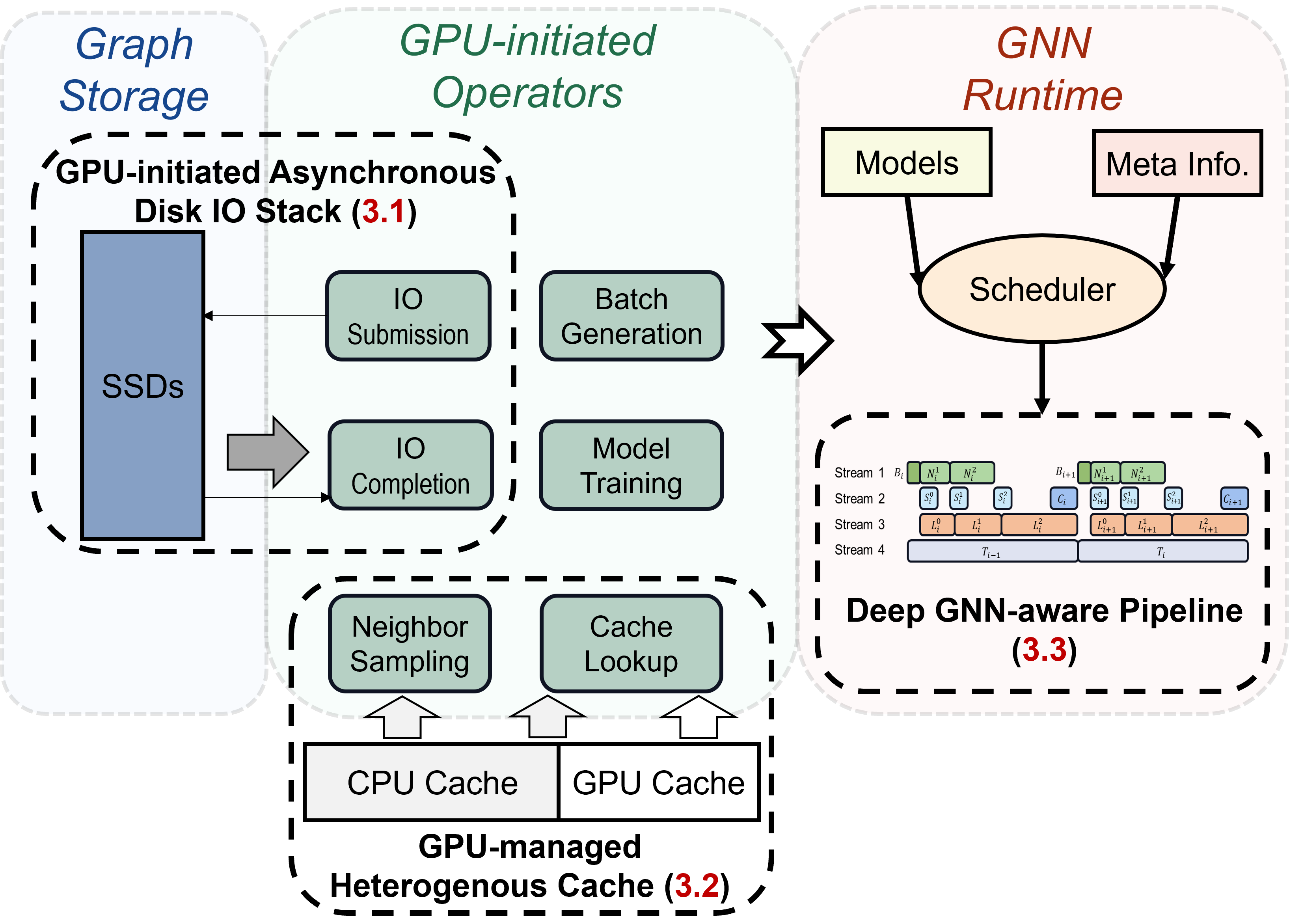} 
    \vspace{-4ex}
	\caption{\SystemName{} Overview.}
    \vspace{-3ex}
	\label{overview}}
\end{figure}

To address the issues mentioned in Section~\ref{motivation_sec}, we propose \SystemName{}, an efficient out-of-core system that can train GNN on terabyte-scale graphs with in-memory performance. The key idea is to minimize the GPU resource overhead for direct SSD accesses, and thus have sufficient GPU resources for heterogeneous cache-enabled in-memory GNN training that fully pipelines with direct SSD accesses and minimizes the number of costly SSD accesses. Figure~\ref{overview} illustrates the overall structure of \SystemName{}. Specifically, we propose three key components in \SystemName{}:

\begin{itemize}[topsep=5pt, leftmargin=*]
\item {\SystemName{} takes multiple SSDs as the graph storage and uses a GPU-initiated asynchronous disk IO stack that only requires about 30\% GPU cores to reach the almost maximal disk IO throughput and wastes no GPU cores between IO submission and IO completion to leave the majority of GPU cores for other GNN kernels (Subsection~\ref{design1});
}

\item{
\SystemName{} builds a GPU-managed heterogeneous cache that extends the cache space on heterogeneous CPU and GPU memory and leverages GPU’s parallelism to boost the cache lookup throughput (Subsection~\ref{design2}); 
}

\item {
\SystemName{} implements a deep GNN-aware pipeline as the GNN runtime that fully utilizes GPU computation cycles, maximizing system throughput (Subsection~\ref{design3}).  
The first two components as well as other GNN operations like model training are implemented as GPU-initiated operators. \SystemName{} schedules and executes all these operators in the deep GNN-aware pipeline. 
}

\end{itemize}

\subsection{GPU-initiated Asynchronous Disk IO Stack}
\label{design1}

To solve the issues I$_1$ and I$_3$, we propose a GPU-initiated asynchronous disk IO stack for feature extraction from disks to limit the number of GPU cores for the IO stack (necessary for issue I$_4$), as shown in Figure~\ref{iostack_ours}. We adopt two decoupled kernels to achieve the goals, namely thread-level parallel IO command submission and asynchronous completion handling. Such a design has two benefits: 1, only requiring 30\% GPU cores to submit sufficient IO requests to reach the almost maximal IO throughput; 2, wasting no GPU cores between IO submission and IO completion leaving the majority of GPU cores for other GNN kernels. In this subsection, we will explain how the two techniques achieve the benefits.

\subsubsection{Thread-level Parallel IO Command Submission} $\\$
To address GIDS's low IO parallelism issue caused by low thread utilization, \SystemName{} proposes a thread-level parallel IO command submission technique that fully utilizes GPU parallelism to maximize the disk IO throughput. 

First, the command submission kernel in \SystemName{} gets the sampled vertices' ID list after neighbor sampling. Second, for each vertex, the command submission kernel generates an NVMe command for extracting the corresponding vertex feature from SSDs. The NVMe command contains the SSD logic block ID and the GPU physical memory address of the temporary IO buffer. Each thread of the kernel is capable of binding multiple commands and sending commands to multiple parallel SQs. 

As such, \SystemName{} can parallelize the command submission at thread level, i.e., the utilization of all threads in each warp can be maximized. Moreover, binding multiple requests in one thread reduces the total thread number to submit enough NVMe commands to maximize the disk IO throughput. As a result, \SystemName{} only needs about 30\% GPU cores (see experiment~\ref{io_stack_exp}) to achieve almost the maximal disk IO throughput.

% Decoupling cache and SSD IO makes command submission efficient and lightweight.

% Besides, \SystemName{} adopts a doorbell batching mechanism. The GPU rings doorbell signals to SSD when all commands are filled to SQs. Each queue rings its doorbell only once, rather than one doorbell per command. This further reduces fragmented PCIe transactions between GPU and SSDs. 

\subsubsection{Asynchronous IO Completion Handling} $\\$
To avoid the large completion handling time, we propose an asynchronous IO completion handling kernel, as shown in Figure~\ref{iostack_ours}. 
In the completion handling kernel, each warp manages multiple parallel CQs, reducing the number of thread blocks of the pooling kernel. One leader thread in each warp polls for the completion entry. After the leader thread gets a completion entry, all threads in the warp move data from the temporary IO buffer to the feature buffer. The latency of polling a completion entry and data movement inside GPU memory is orders of magnitudes less than disk IO. Thus the completion handling kernel can bring negligible overhead on the overall throughput. Due to the multiple parallel CQs design, kernels in IO completion handling also just need about 30\% GPU cores to maximize the throughput. 

This kernel decouples the IO completion handling from the IO submission, such that massive IO stack threads are not active for waiting IO completion. Compared to synchronous IO stack like BaM~\cite{qureshi2023gpu}, this mechanism frees up GPU computation resources during the pooling stage, and thus leaves the majority of GPU cores for other GNN kernels. %thus enabling efficient overlap between the IO stack and other kernels.

\begin{figure}[t]
\vspace{-0ex}
	\centering
	{\includegraphics[width=\linewidth]{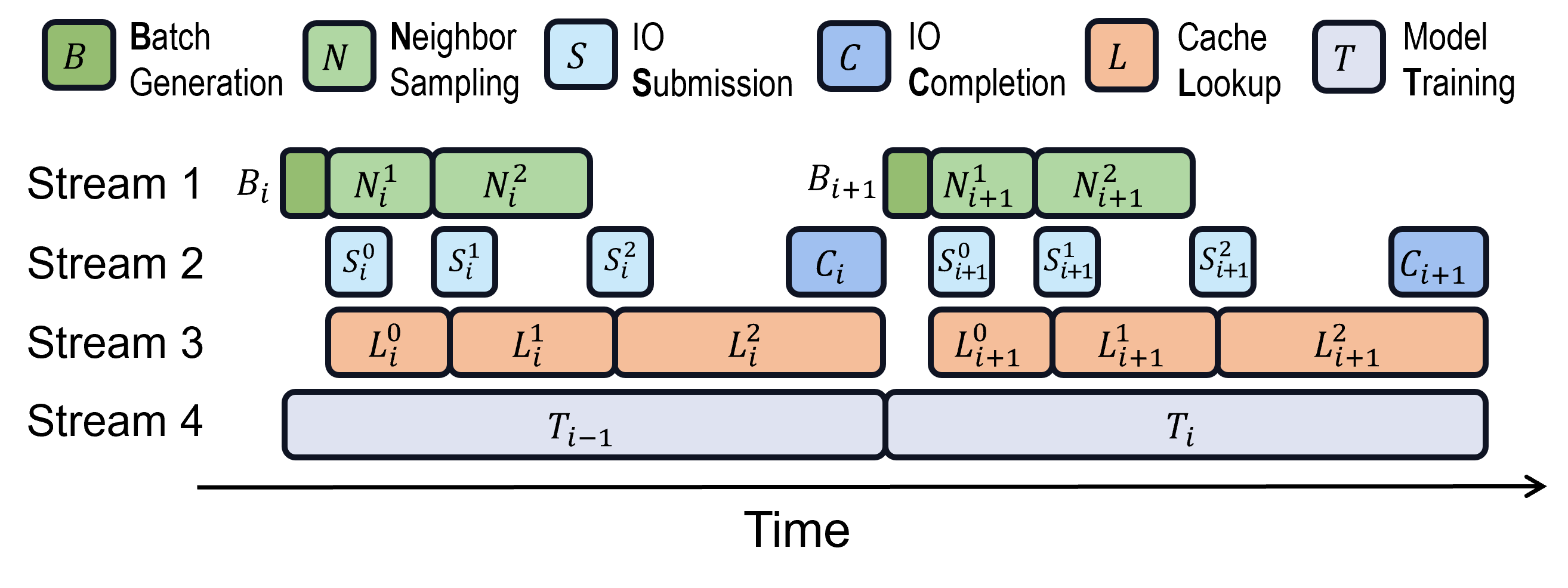} 
    \vspace{-4ex}
	\caption{An example of \SystemName{}'s deep GNN-aware pipeline for a two-hop GraphSAGE~\cite{hamilton2017inductive}. $i$ represents the mini-batch ID. The superscripts stand for the current number of hops for the operator.}
    \vspace{-3ex}
	\label{pipeline_design}}
\end{figure}

\subsection{GPU-managed Heterogeneous Cache}
\label{design2}
To overcome the issues I$_2$ and I$_5$ of existing systems, we propose a GPU-managed heterogeneous cache, as shown in Figure~\ref{overview}. The cache utilizes heterogeneous CPU and GPU memory as the physical cache space and takes an effective static pre-sampling-based cache policy~\cite{yang2022gnnlab, sun2023legion}. We utilize GPU to initialize cache items and make use of GPU's massive parallelism to enable a high throughput cache lookup. With this design, \SystemName{} further augments the IO throughput to achieve an in-memory-like training performance, even when the graphs reach terabyte-scale (See~\ref{cache_exp_section}).

\subsubsection{Cache Placement Strategy}$\\$
The heterogeneous cache consists of two parts: the GPU cache and the CPU cache. 

\noindent{\textbf{GPU Cache. }
The GPU cache is physically stored in the GPU's device memory, thus offering nearly 2TB/s bandwidth on an A100 GPU. According to the physical nature of the GPU cache, we place the most frequently accessed feature data in the GPU cache. As such, this design could mitigate the data transfer volume from CPU memory and SSDs.}

\noindent{\textbf{CPU Cache. }
The CPU cache is physically stored in the CPU's DRAM and connected to the GPU by PCIe, which offers nearly 20 GiB/s bandwidth (PCIe 4.0x16). We take the CPU memory as the CPU cache for two reasons. First, the CPU memory has much lower latency and higher throughput under finer-granularity data access compared to SSDs\footnote{The smallest data access granularity of a typical NVMe SSD is 512 Bytes.}, thus compensating for underutilized PCIe bandwidth when accessing data from SSDs. Second, the CPU cache offers a much larger cache space compared to the GPU cache. For most graphs, the topology data size is far smaller than the feature sizes (See Table~\ref{dataset}). Moreover, the real-world graph data is often skewed, e.g., caching only 10\% of data in the CL dataset (See Table~\ref{dataset}) can reduce 70\% of data transfer volume from SSDs. 
We observe that the capacity of the currently mainstream CPU DRAM (up to 1TB) is large enough to obviously reduce SSD accesses. As such, we store the entire topology data in the CPU cache to avoid byte-grained data access from SSDs and fill the rest part of the CPU cache space with the second-hottest features. 
% Figure~\ref{cache_exp_section} shows \SystemName{} can achieve an in-memory-like training throughput, even when the graphs reach terabyte-scale.
}

\subsubsection{Cache Management by GPU}$\\$
\SystemName{} relies on GPU to initialize the heterogeneous cache and issues cache lookups. 

\noindent{\textbf{Cache Initialization. } Before the GNN training starts, \SystemName{} first loads the entire topology data to the CPU cache.\footnote{Topology size is always lower than the CPU memory capacity. In future work, we generalize \SystemName{} to out-of-memory topology. } Second, \SystemName{} relies on GPU to run an epoch of pre-sampling~\cite{yang2022gnnlab, sun2023legion} and collects all vertices' hotness. Subsequently, \SystemName{} uses GPU to sort all vertices by their hotness in descending order. According to our cache placement strategy, \SystemName{} selects the hottest features to fill up the GPU cache and the second-hottest features to fill up the CPU cache. 
}

\noindent{\textbf{Cache Lookup.} \SystemName{} launches GPU kernels for neighbor sampling and feature cache lookup. Inside these kernels, GPU threads directly access the cached data in CPU memory by UVA~\cite{uvaref} technique or in GPU memory. As such, multiple threads concurrently look up the cache, thus efficiently hiding the latency of memory access. We carefully choose an optimal number of warps for the neighbor sampling and cache lookup kernel so that it maximizes the throughput while limiting the usage of GPU computation resources.
}

\begin{figure*}[t]
    \subfigure[GraphSAGE]{
        \label{fig_endtoend_sage}
        \includegraphics[width=0.48\linewidth]{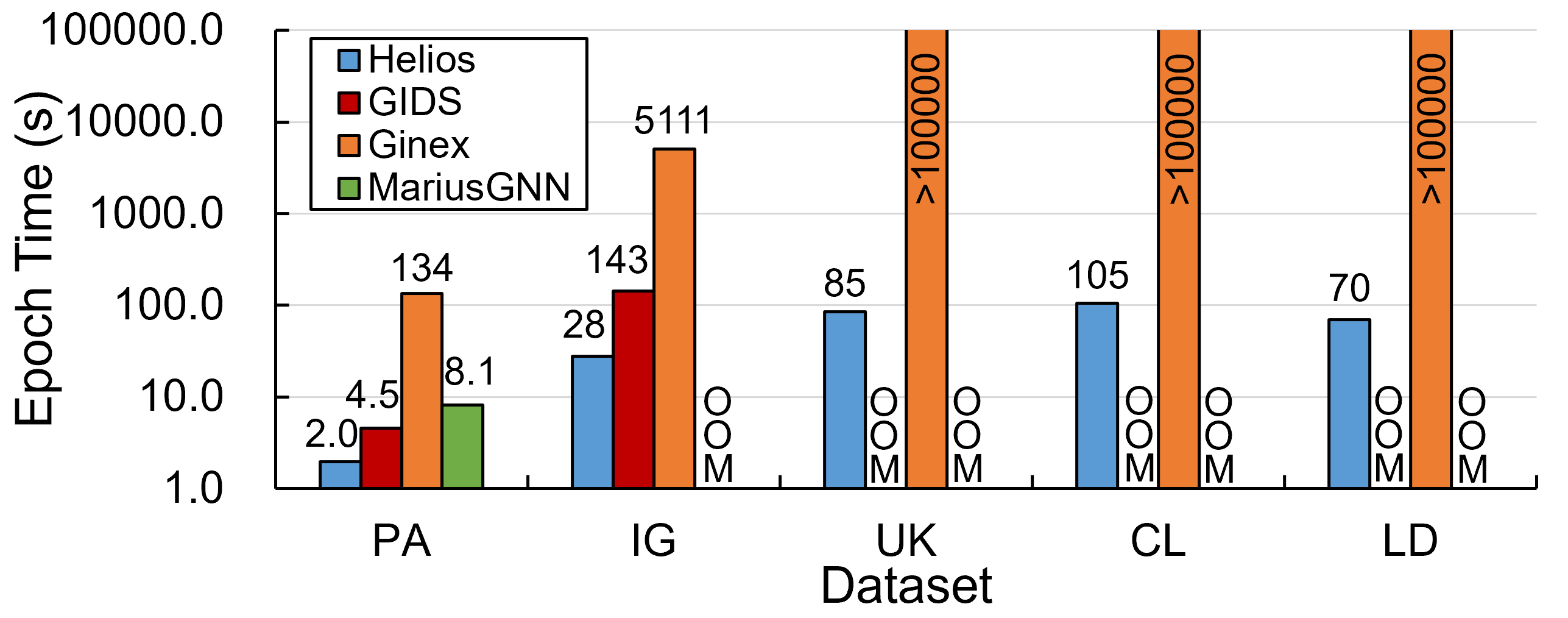}
    }
    \hfill    
    \subfigure[GCN]{
        \label{fig_endtoend_gcn}
        \includegraphics[width=0.48\linewidth]{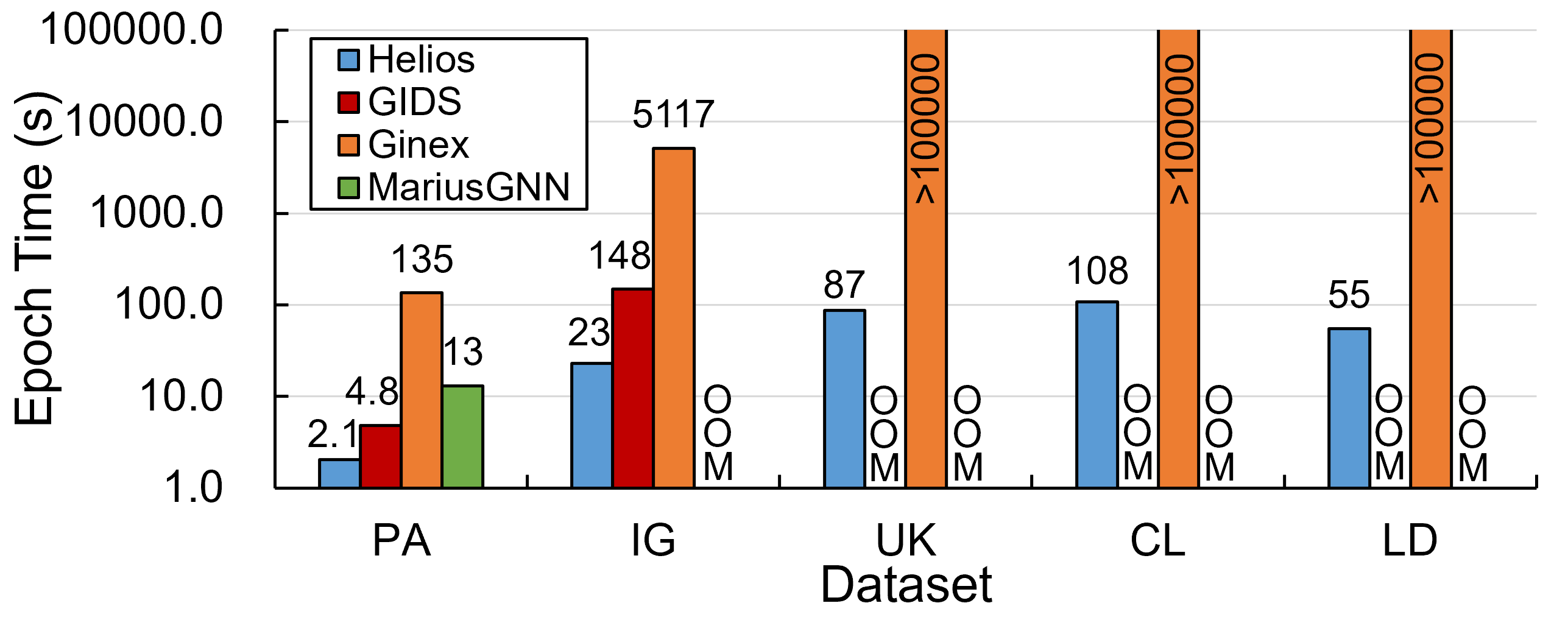}
    }
\vspace{-4ex}   
\caption{End-to-end performance of \SystemName{} and the baselines on various datasets.} 
   \vspace{-3ex}
    \label{fig_endtoend} 
\end{figure*} 

\subsection{Deep GNN-aware Pipeline}
\label{design3}
To solve the issue I$_4$, we propose a deep GNN-aware pipeline to boost the overall throughput of \SystemName{}. Figure~\ref{pipeline_design} illustrates an example of a 2-hop GraphSAGE model's pipeline execution. There are three steps in \SystemName{} to execute the deep GNN-aware pipeline.

\noindent{\textbf{Pipeline Execution Plan Construction.}
Given the GNN models' specification, the scheduler in \SystemName{} first decomposes GNN model execution into a sequence of GPU-initiated operators: IO submission and IO completion (See Subsection~\ref{design1}); neighbor sampling and cache lookup (See Subsection~\ref{design2}); batch generation and model training. For example, in Figure~\ref{pipeline_design}, $S_i^0$ represents the operator that submits the feature extraction requests to disks for the seed vertices of the $i$-th mini-batch. $C_i$ means IO completion handling of all sampled vertices' features in the $i$-th mini-batch. Second, \SystemName{} analyzes the two-level data dependency of the GNN model execution: sample-train dependency and multi-hop dependency. Accordingly, \SystemName{} generates a two-level pipeline execution plan including an inter-mini-batch pipeline and an intra-mini-batch pipeline that will concurrently run GPU-initiated operators with no dependency.
}

\noindent{\textbf{Operator Resources Allocation.}
In this step, \SystemName{} gets the meta information including the number of GPU SMs and GPU memory size, and allocates optimal resources for each operator. 
For the intra-mini-batch pipeline, \SystemName{} allocates specific GPU SMs by setting the GPU kernels' grid size and block size. For example, \SystemName{} sets the number of thread blocks of IO submission and IO completion operators to 32 in our hardware setting as discussed in Subsection~\ref{design1}. For the inter-mini-batch pipeline, \SystemName{} decouples the mini-batch preparation and model training into two separate processes and limits the SM usage of training operators by CUDA-MPS~\cite{mpsref}. 
}

\noindent{\textbf{Pipelined Execution.}
\SystemName{} launches all operators with specific GPU resources to different asynchronous streams according to the computation plan. As shown in Figure~\ref{pipeline_design}, with the intra-mini-batch pipeline design, neighbor sampling, IO submission, IO completion, and cache lookup operators can be well overlapped with each other. As such, \SystemName{} can maximize the utilization of GPU PCIe bandwidth. Moreover, with the inter-mini-batch pipeline, the IO from all memory hierarchies can be well overlapped with the computation of GNN model training. 
}

%% file: context/evaluation.tex
\section{Evaluation}
\subsection{Experimental Setting}

\begin{table} [t]
\renewcommand\arraystretch{1.4}
\setlength\tabcolsep{8pt}
	\centering

	%\begin{spacing}{0.3}
		\begin{scriptsize}
		
	\vspace{-1ex}
	\caption{Dataset Statistics} \vspace{-1.5ex}
	\label{dataset}
	\begin{tabular}{|c||c|c|c|c|c|c|c|}
		\hline
		{\bf Dataset} &  {\bf PA}&  {\bf IG }  &  {\bf UK} &  {\bf CL} & {\bf LD} \\
		\hline
            \hline
		{\bf Vertices Num.} &111M & 269M & 0.79B & 1B & 5.6B \\ 
		\hline
      	{\bf Edges Num.} & 1.6B & 4B & 47.2B & 42.5B & 10B \\ 
		\hline
            {\bf Topology Size} & 14GB & 34GB & 384GB & 348GB & 125GB \\ 
		\hline
  		{\bf Feature Dim.} & 128 & 1024 & 1024 & 1024 & 1024\\ 
		\hline
    	{\bf Feature Size} & 56GB & 1.1TB & 3.2TB & 4.1TB & 23TB \\ 
		\hline
	\end{tabular}

		\end{scriptsize}
	\vspace{-4ex}
	%\end{spacing}
\end{table}

\noindent{\bf Experimental Platform.} The experiments are conducted using a single server with one NVIDIA 80GB-PCIe-A100 GPU and 12 $\times$ 3.84TB Intel P5510 NVMe SSDs. GPU and SSDs are connected by PCIe 4.0x16. Additionally, the server has 2 $\times$ Intel(R) Xeon(R) Gold 5320 CPU (2 $\times$ 52 threads) @ 2.20GHz and 768 GB CPU memory.

\noindent{\bf GNN Models.} We use two sampling-based GNN models: GraphSAGE~\cite{hamilton2017inductive} and GCN~\cite{kipf2016semi}. Both adopt a 2-hop random neighbor sampling. The sampling fan-outs are 25 and 10. The dimension of the hidden layers in both models is set to 256. Similar to existing work~\cite{yang2022gnnlab, sun2023legion}, the batch size is set to 8000. Unless explicitly explained, node classification is used as the GNN task.

\noindent{\bf Datasets. } We conduct our experiments on multiple real-world graph datasets with various scales. Table~\ref{dataset} shows the dataset characteristics. The Paper100M (PA) is available in Open Graph Benchmark~\cite{hu2020open}. The IGB-HOM (IG) is from the IGB dataset~\cite{khatua2023igb}. The UK-2014 (UK), and Clue-web (CL) are from WebGraph~\cite{BoVWFI, BRSLLP, BCSU3, BMSB}. The LDBC-SNB-Bi-SF3000 (LD) is available at the LDBC social network benchmarks~\cite{angles2020ldbc}. Because UK, CL, and LD have no feature, we manually generate the features with the dimension specified as $1024$, following IGB's setting. Similar to PA's setting, we choose 1\% of vertices from each graph as training vertices. 

\noindent{\bf Baselines. } We use the state-of-the-art disk-based GNN systems, Ginex~\cite{park2022ginex}, MariusGNN~\cite{waleffe2023mariusgnn} and GIDS~\cite{park2023accelerating} as the baseline systems. For CPU-managed baselines Ginex and MariusGNN, we set the number of CPU threads to 96. 

\begin{figure}[t]
    \begin{center}
        \subfigure[\label{fig_in_memory_sage} GraphSAGE]{
            \includegraphics[width=0.47\linewidth]{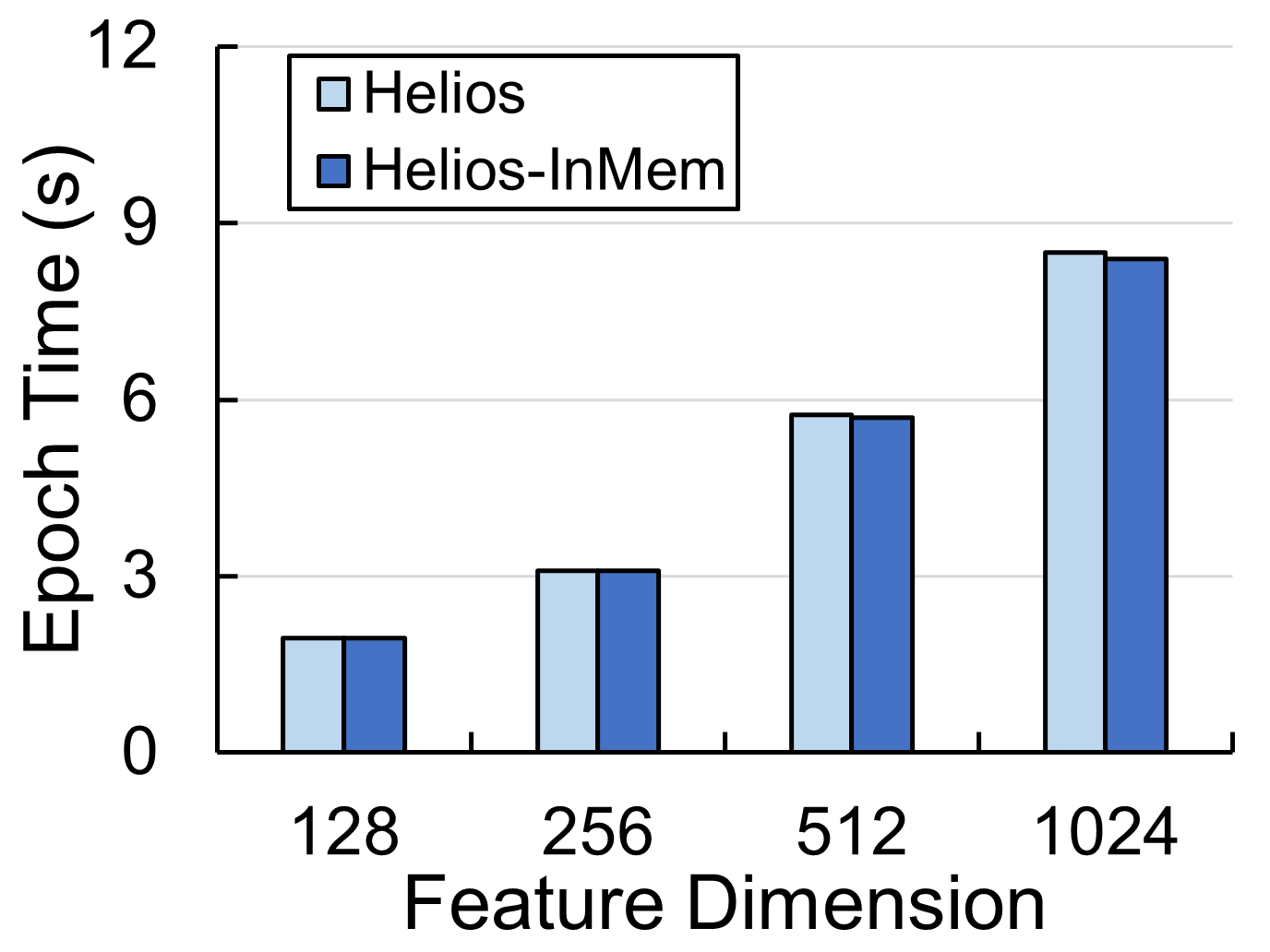}
        }
        \hfill
        \subfigure[\label{fig_in_memory_gcn} GCN]{
            \includegraphics[width=0.47\linewidth]{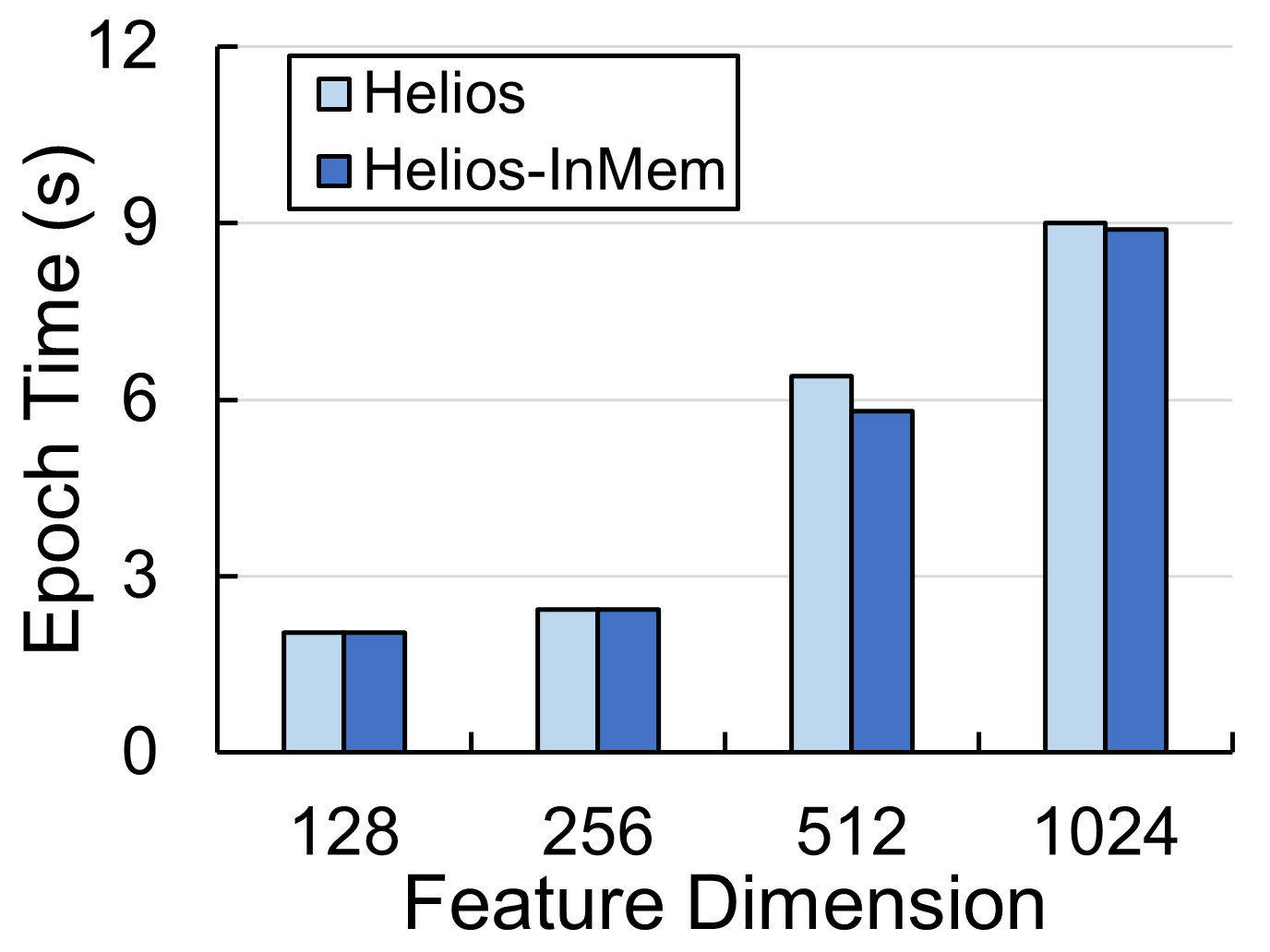}
        }
    \end{center}
	\vspace{-4ex}
    \caption{\label{fig_in_memory} \SystemName{}'s performance compared with in-memory system.}
	\vspace{-4ex}
\end{figure}

\subsection{End-to-end Performance}

To demonstrate the efficiency of \SystemName{}, we first compare \SystemName{} with all three baselines~\cite{park2022ginex, waleffe2023mariusgnn, park2023accelerating}. Meanwhile, we compare \SystemName{} with \SystemName{}-InMem to show that \SystemName{} can match the training performance with the corresponding in-memory systems.

\noindent{\textbf{Comparing with State-of-the-art Baselines.} We compare \SystemName{} with Ginex~\cite{park2022ginex}, MariusGNN~\cite{waleffe2023mariusgnn}, and GIDS~\cite{park2023accelerating} on all datasets in Table~\ref{dataset} and two GNN models. And we use 12 SSDs to store the datasets. For all systems, we store the entire topology data in CPU memory for fair comparison.

Figure~\ref{fig_endtoend} shows the average epoch time of each system on terabyte-scale datasets (IG, UK, CL, LD). We observe that \SystemName{} outperforms GPU-managed baseline GIDS by up to 6.43$\times$. However, on larger UK, CL, and LD datasets, GIDS runs out of GPU memory due to over 80GB metadata in page cache design of BaM~\cite{qureshi2023gpu}. 
Compared to CPU-managed baseline Ginex, \SystemName{} achieves over 182$\times$ speedup. On these datasets, MariusGNN runs out of memory due to large memory consumption during preprocessing.

On the smallest dataset PA, all systems store all topology and features in CPU memory. In this case, \SystemName{} outperforms GIDS, MariusGNN, and Ginex by up to 2.3$\times$, 6.2$\times$, and 30$\times$, respectively, indicating \SystemName{} still has good performance on graphs that can be fit in memory.
}

\noindent{\textbf{Comparing with In-memory Systems. }
To compare \SystemName{} with \SystemName{}-InMem, we use the PA dataset and two GNN models for evaluation. To show the scalability of the dataset, we modify the feature dimension from the original 128 to 1024. For both systems, we fill all the available GPU memory by feature cache. For \SystemName{}, we only cache 10\% data in CPU memory while for \SystemName{}-InMem, we store all the feature data in CPU memory.

Figure~\ref{fig_in_memory} illustrates the average epoch time of \SystemName{} compared to \SystemName{}-InMem. We can see that with both GNN models, \SystemName{} can achieve 91\%\textasciitilde 99\% throughput of \SystemName{}-InMem, indicating that even with out-of-core training, \SystemName{} can still match the training throughput with in-memory systems. 
}

\begin{figure}[t]
    \begin{center}
        \subfigure[\label{fig_io_ssdnum} IO Throughput w.r.t SSD number with feature dimension of 1024]{
            \includegraphics[width=0.47\linewidth]{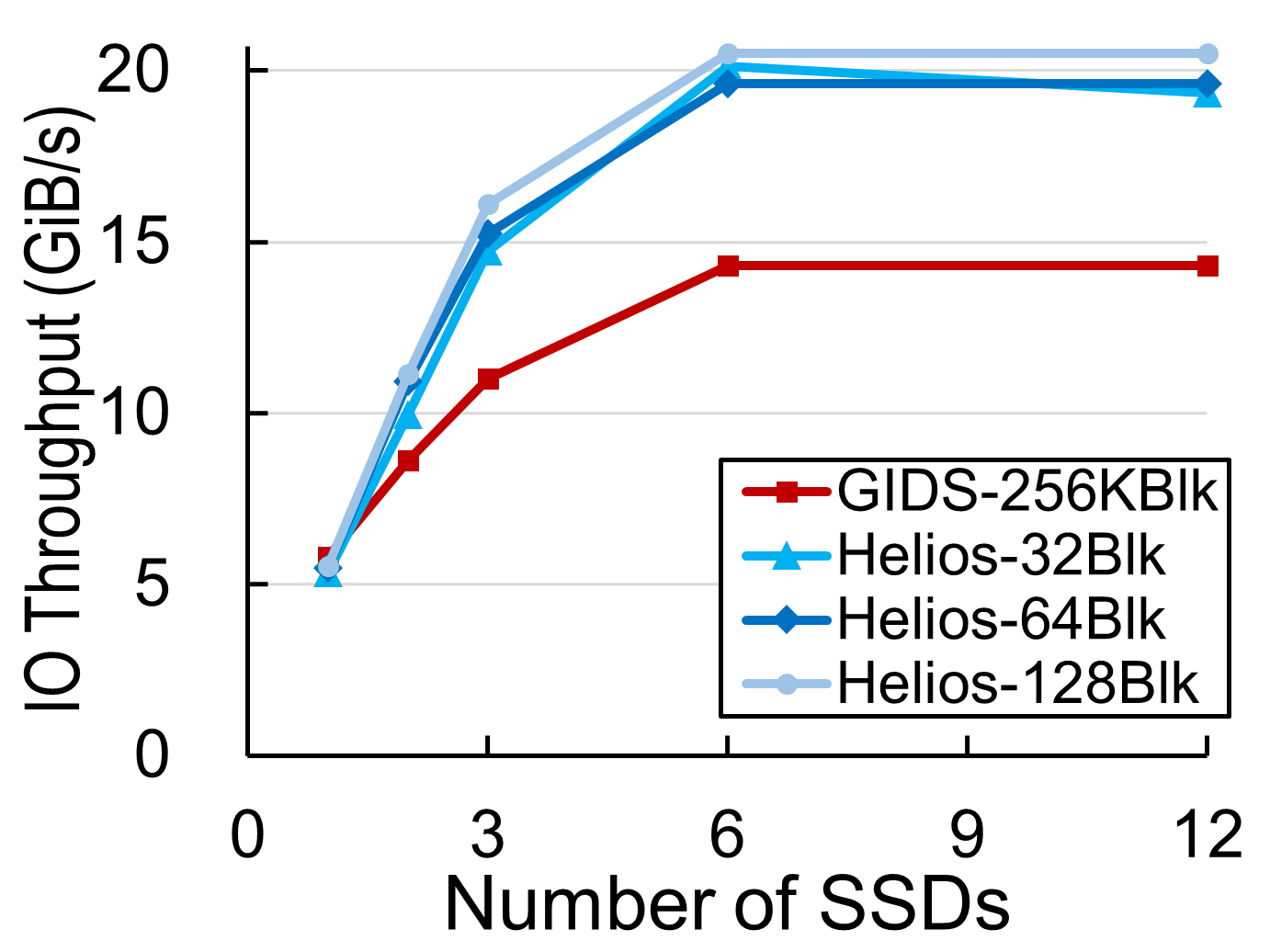}
        }
        \hfill
        \subfigure[\label{fig_io_featdim} IO Throughput w.r.t feature dimension with 12 SSDs]{
            \includegraphics[width=0.47\linewidth]{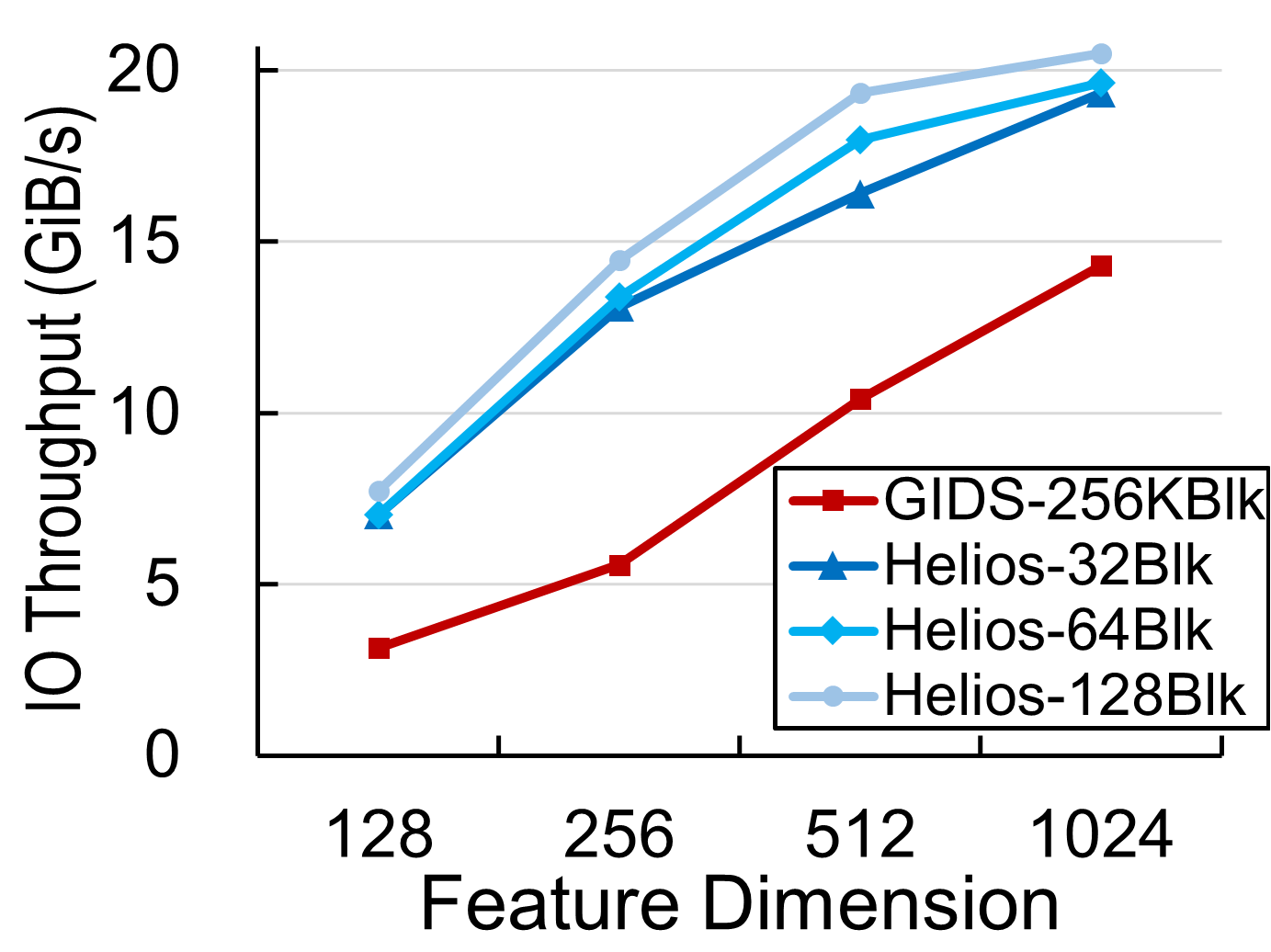}
        }
    \end{center}
    \vspace{-3ex}
    \caption{\label{fig_io} Impact of GPU-initiated asynchronous disk IO stack.}
    \vspace{-4ex}
\end{figure}

\subsection{Impact of GPU-initiated Asynchronous Disk IO Stack}
\label{io_stack_exp}

In this experiment, we evaluate the effect of GPU-initiated asynchronous disk IO stack. The goal is to examine the achieved disk IO throughput as well as the used ratio of GPU computation resources of our design.

We compare the IO stack of \SystemName{} and GIDS~\cite{park2023accelerating}. GIDS exhausts all GPU cores for IO kernels (256K thread blocks, 100\% GPU core utilization). For \SystemName{}, we set the GPU thread block number to 32, 64, and 128 for kernels in \SystemName{}'s GPU-initiated asynchronous disk IO stack, which is about 30\%, 60\% and 100\% GPU core utilization respectively, as shown in Figure~\ref{fig_io}. For example, \SystemName{}-32Blk represents the GPU-initiated asynchronous disk IO stack with 32 thread blocks in each kernel. 
We fix the feature dimension to 1024 and vary the SSD numbers from 1 to 12 (Figure~\ref{fig_io_ssdnum}). Besides, we fix the SSD number to 12 and vary the feature dimension from 128 to 1024 (Figure~\ref{fig_io_featdim}). 

Figure~\ref{fig_io} shows that \SystemName{}'s IO stack outperforms GIDS's IO stack under different numbers of SSDs and feature dimensions, because the decoupled IO stack design can submit sufficient parallel IO requests to efficiently maximize the disk IO throughput, as discussed in Subsection~\ref{design1}. As shown in Figure~\ref{fig_io_ssdnum}, \SystemName{} can reach the maximal IO throughput under PCIe 4.0x16 with 6 SSDs, which means \SystemName{} can achieve the similar throughput of the in-memory system, even when the graphs grow into terabyte-scale.

Meanwhile, \SystemName{} only needs 32 thread blocks for IO stack kernels to achieve a comparable IO throughput with GIDS using 128 thread blocks (all GPU cores). Such a benefit allows \SystemName{} to have the majority of GPU cores to perform computation, rather than to wait for the completion of IO commands. %leaves a large room for \SystemName{} to overlap communication with computation. 

\subsection{Impact of GPU-managed Heterogeneous Cache}
\label{cache_exp_section}

\begin{figure}[t]
    \begin{center}
        \subfigure[\label{fig_cache_ssdnum_time} Epoch Time]{
            \includegraphics[width=0.47\linewidth]{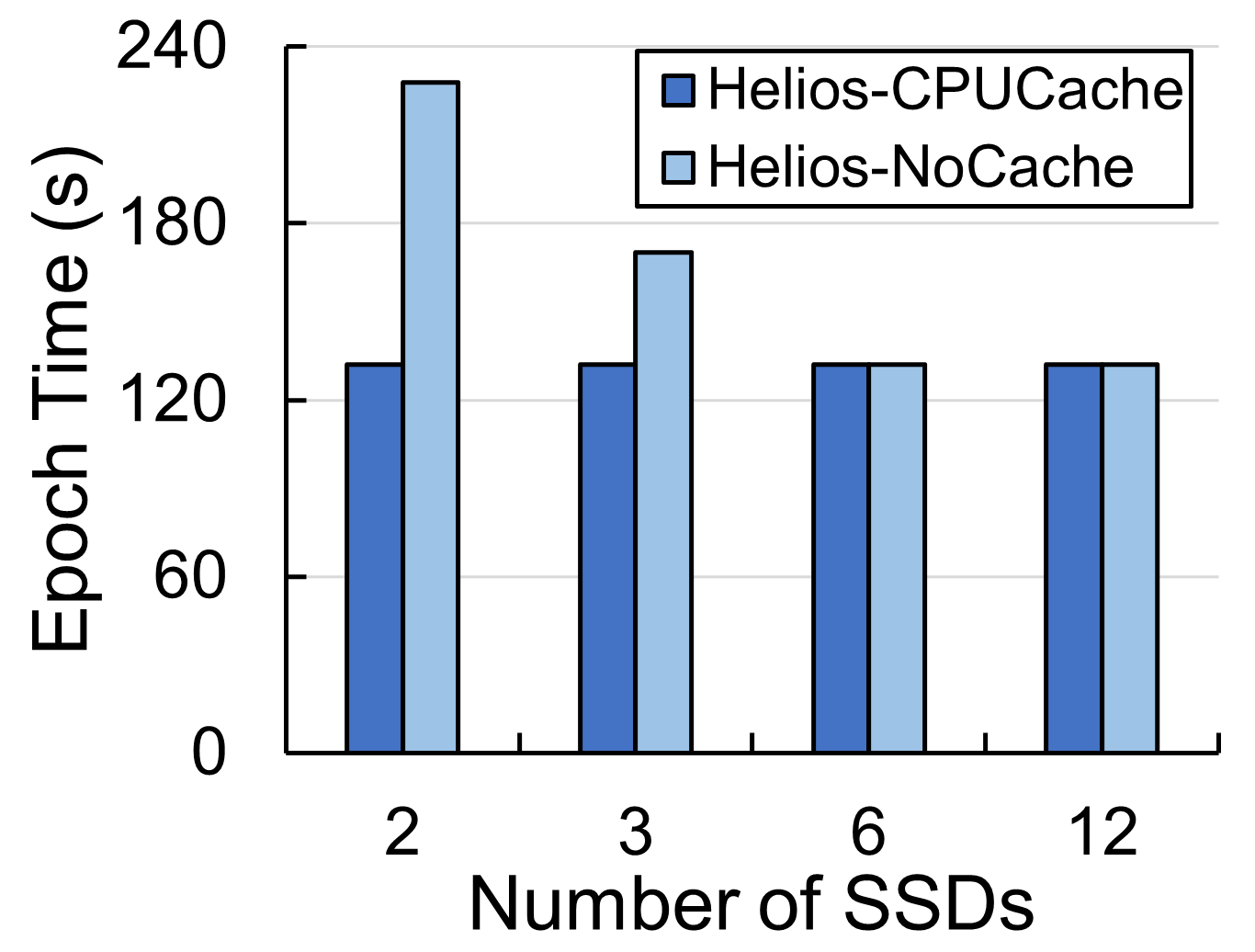}
        }
        \hfill
        \subfigure[\label{fig_cache_ssdnum_bw} GPU PCIe Throughput]{
            \includegraphics[width=0.47\linewidth]{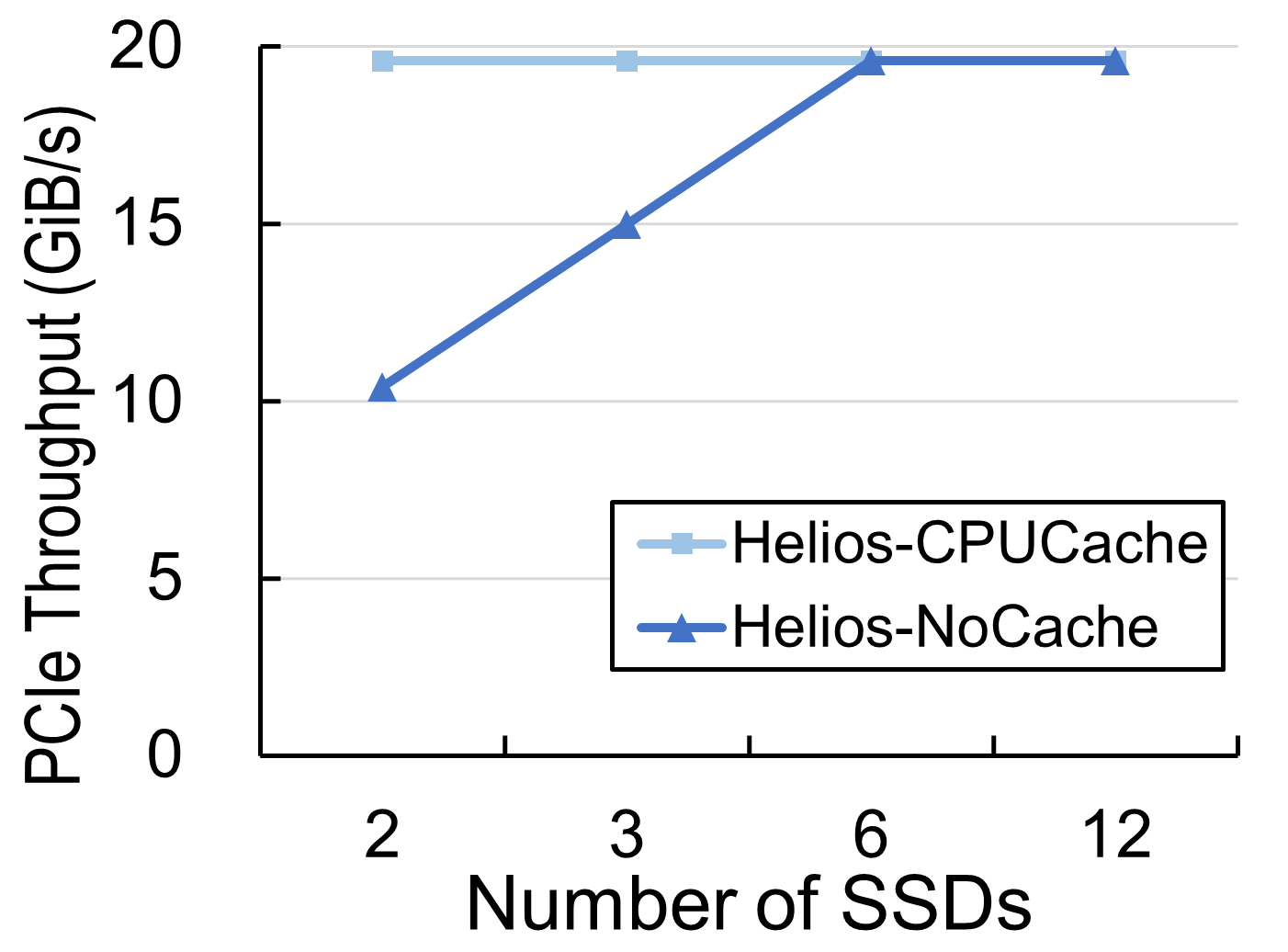}
        }
    \end{center}
    \vspace{-4ex}
    \caption{\label{fig_cache_ssdnum} Impact of CPU cache on different SSD numbers.}
    \vspace{-4ex}
\end{figure}

\begin{figure}[t]
    \begin{center}
        \subfigure[\label{fig_cache_featdim_time} Epoch Time]{
            \includegraphics[width=0.47\linewidth]{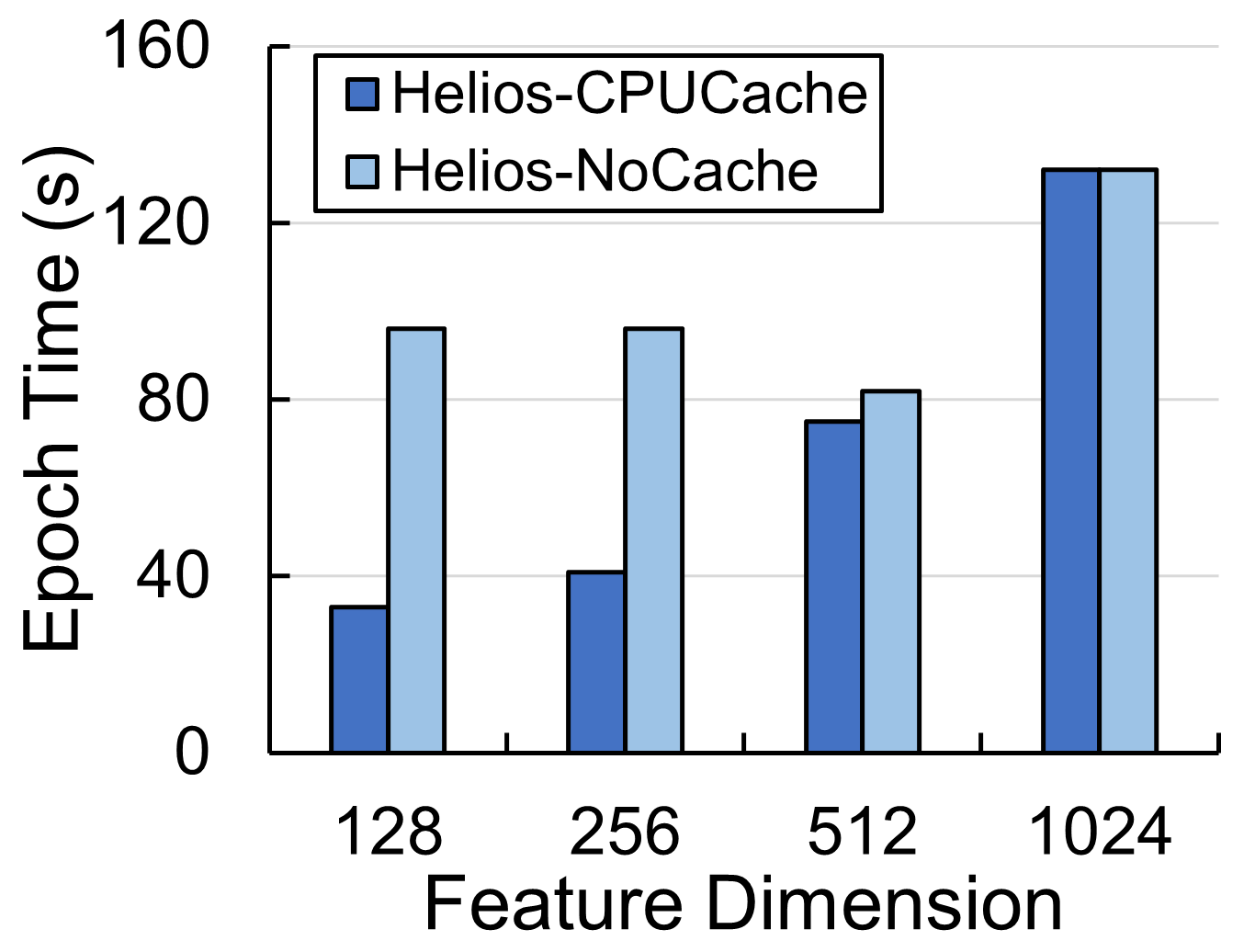}
        }
        \hfill
        \subfigure[\label{fig_cache_featdim_bw} GPU PCIe Throughput]{
            \includegraphics[width=0.47\linewidth]{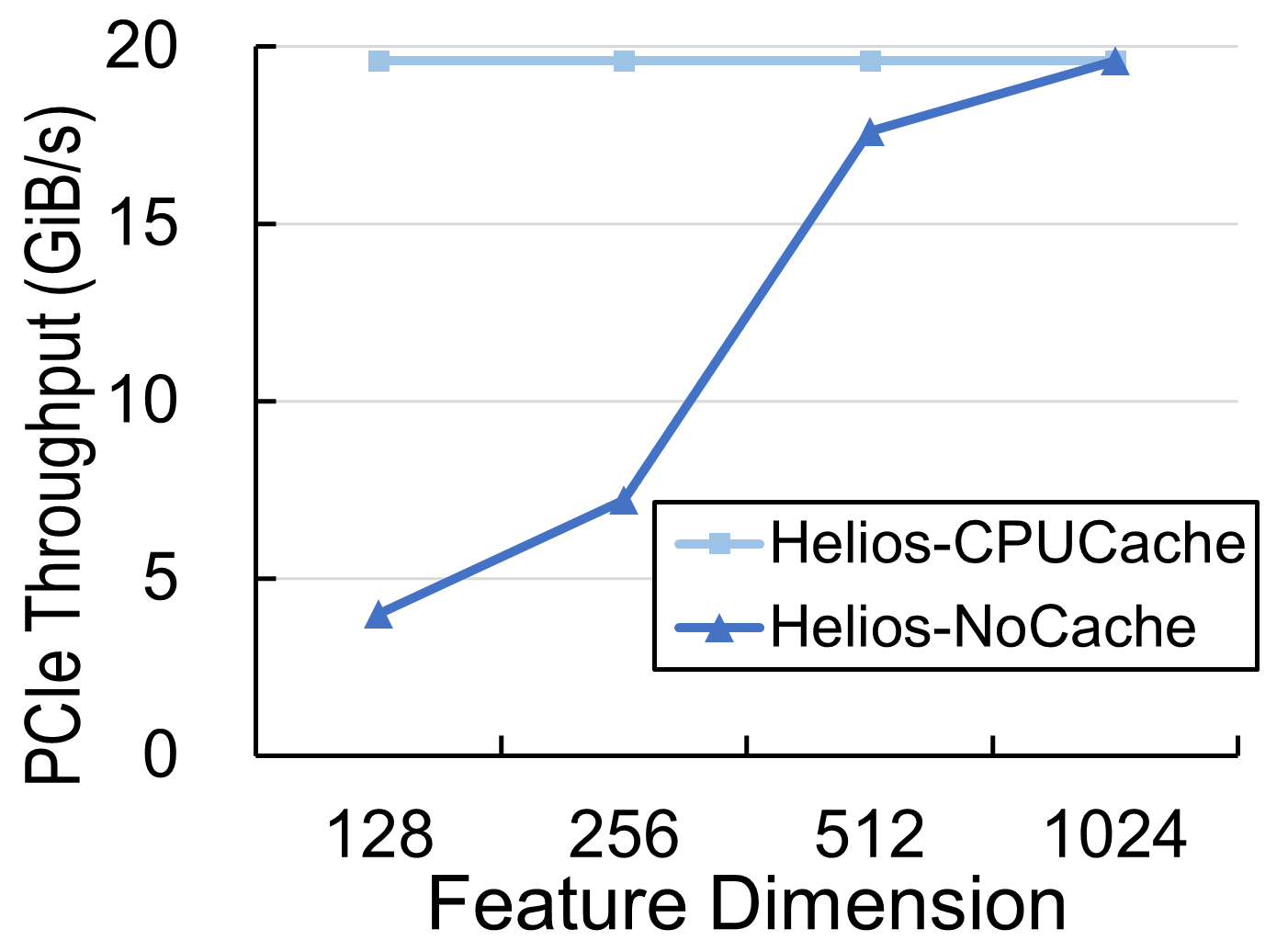}
        }
    \end{center}
    \vspace{-4ex}
    \caption{\label{fig_cache_featdim} Impact of CPU cache on different feature dimensions.}
    \vspace{-4ex}
\end{figure}

\begin{figure}[t]
\vspace{-0ex}
	\centering
	{\includegraphics[width=0.65\linewidth]{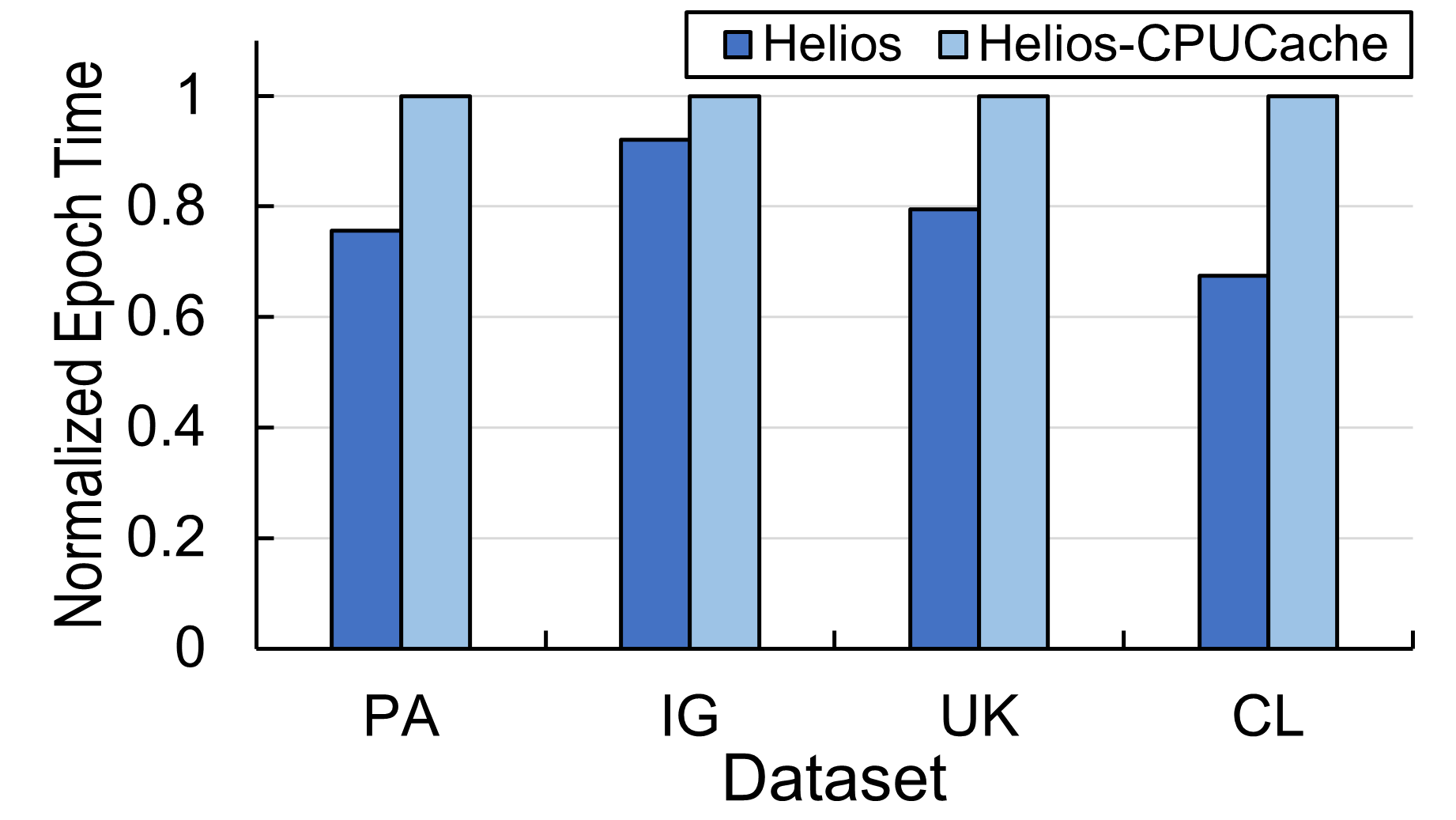}} 
%    \vspace{-1ex}
  \vspace{-2ex}
	\caption{Impact of GPU Cache.}
	\label{fig_cache_dataset}
 \vspace{-3ex}
\end{figure}

In this experiment, we evaluate the impact of GPU-managed heterogeneous cache in \SystemName{}. 

\noindent{\textbf{The Impact of CPU Cache.}
We show the impact of CPU cache in \SystemName{} in Figure~\ref{fig_cache_ssdnum} and~\ref{fig_cache_featdim}. To illustrate the impact, we propose two implementations of \SystemName{}, \SystemName{}-CPUCache and \SystemName{}-NoCache. \SystemName{}-CPUCache reads features from both the CPU cache and SSDs, while \SystemName{}-NoCache reads features only from SSDs. We disable GPU caches and maintain CPU cache for graph topology in both implementations. We show the evaluation result on a terabyte-scale CL dataset (4.1TB). 

First, we conduct an experiment with varying SSD numbers from 2 to 12, as shown in Figure~\ref{fig_cache_ssdnum}. The feature dimension of CL is 1024. We can see from Figure~\ref{fig_cache_ssdnum_time} that \SystemName{}-CPUCache outperforms \SystemName{}-NoCache by up to 1.73$\times$ and maintains a stable throughput even with only two SSDs. We can refer to Figure~\ref{fig_cache_ssdnum_bw} for underlying reasons. It shows that \SystemName{}-CPUCache reaches the maximal PCIe
throughput under different numbers of SSDs. This also indicates that \SystemName{} can achieve a throughput comparable with that of the in-memory system, even on terabyte-scale graphs.

Second, we vary the feature dimension of CL from 128 to 1024 (Figure~\ref{fig_cache_featdim}) to show the benefits of CPU cache under different access granularity. Here we use 12 SSDs for evaluation. As shown in Figure~\ref{fig_cache_featdim_time}, \SystemName{}-CPUCache outperforms \SystemName{}-NoCache by up to 2.91$\times$. This is because small access granularity leads to severe low throughput issues of SSDs. We can see from Figure~\ref{fig_cache_featdim_bw} that \SystemName{}-CPUCache maintains a maximal PCIe throughput under all feature dimensions while \SystemName{}-NoCache only achieves down to 20\% PCIe throughput. Again, the results also indicate \SystemName{} can achieve throughput at memory efficiency under different granularities.
}

\noindent{\textbf{The Impact of GPU Cache.}
In this part, we examine the impact of GPU cache on different datasets. Figure~\ref{fig_cache_dataset} illustrates that \SystemName{} with the GPU cache and CPU cache (\SystemName{}) can achieve speedup from 1.09\textasciitilde 1.48$\times$, compared with \SystemName{} with only CPU cache and no GPU cache (\SystemName{}-CPUCache). 
}

\begin{figure}[t]
    \begin{center}
        \subfigure[\label{fig_pipeline_sage} GraphSAGE]{
            \includegraphics[width=0.47\linewidth]{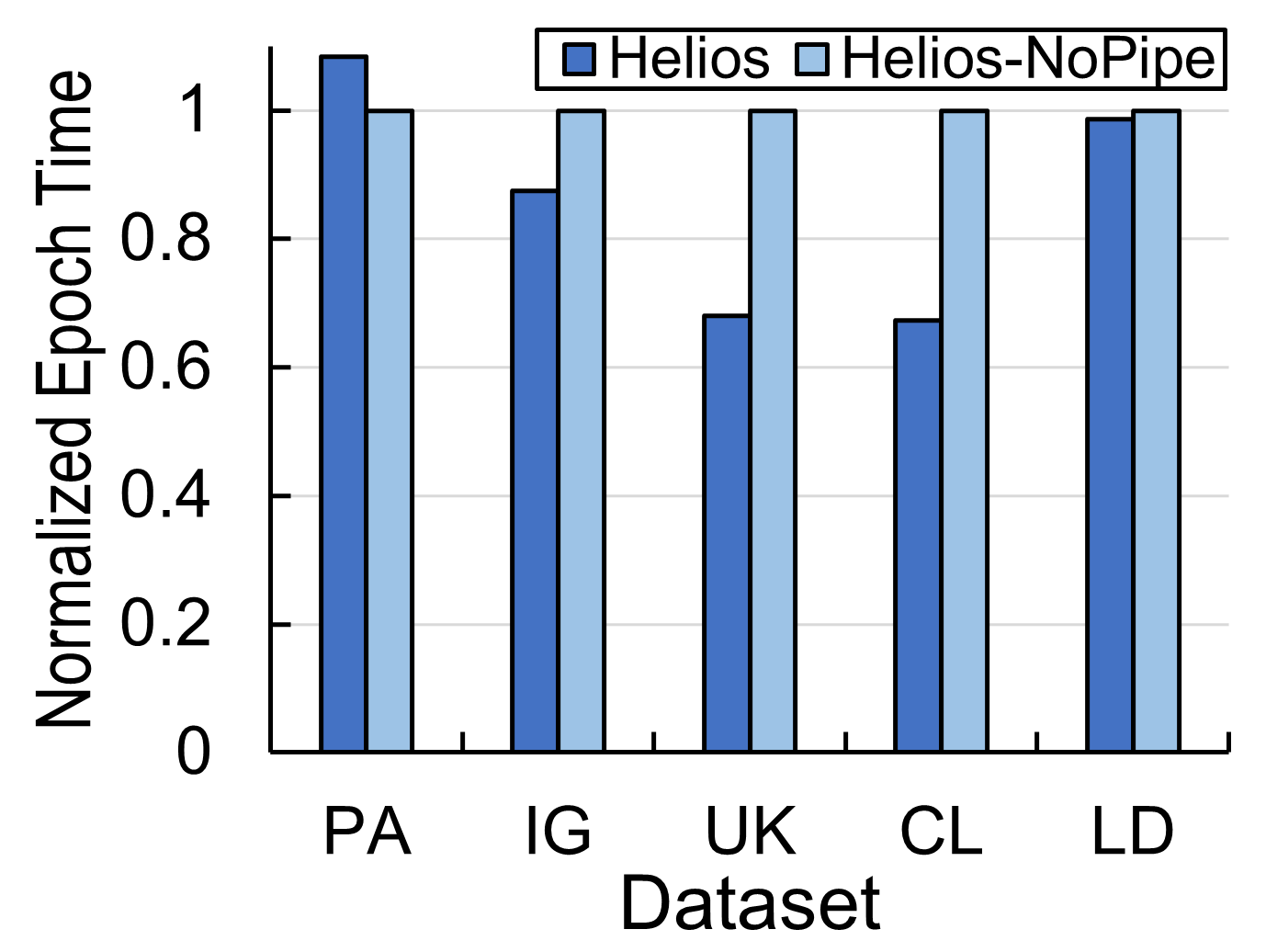}
        }
        \hfill
        \subfigure[\label{fig_pipeline_gcn} GCN]{
            \includegraphics[width=0.47\linewidth]{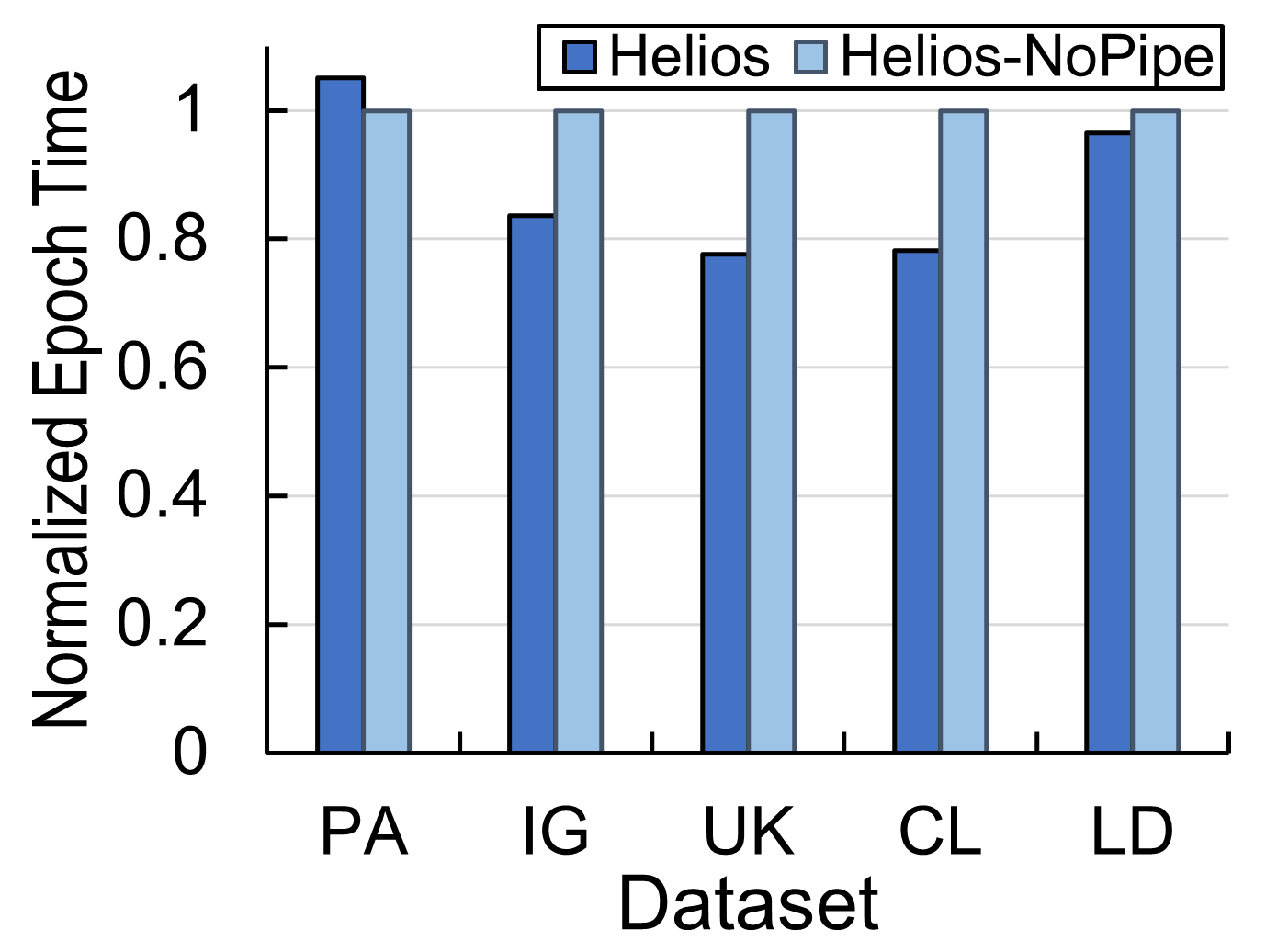}
        }
    \end{center}
    \vspace{-4ex}
    \caption{\label{fig_pipeline} Impact of deep GNN-aware pipeline.}
    \vspace{-3ex}
\end{figure}

\subsection{Impact of Deep GNN-aware Pipeline}
To validate the effectiveness of the deep GNN-aware pipeline (See Subsection~\ref{design3}), we propose \SystemName{}-NoPipe, an implementation
that launches all GPU cores for all stages in the GNN training process and executes all GPU-initiated operators in a serial order. Then we compare \SystemName{} with \SystemName{}-NoPipe. Figure~\ref{fig_pipeline} illustrates the comparison results under two GNN models and all datasets.

\SystemName{} outperforms \SystemName{}-NoPipe by up to 1.49$\times$ for GraphSAGE and 1.29$\times$ for GCN. The speedup mainly comes from the efficient overlapping of computation and data access. However, on the PA dataset, there is a slight performance loss of \SystemName{} compared to \SystemName{}-NoPipe. This is because over 50\% of the PA's features can be cached in GPU memory, leading to almost 99\% GPU cache hit. As a result, the data access time becomes much lower than the model training time, thus the benefit of overlapping data access and model training is slightly lower than the overhead introduced by limiting the training kernels' GPU cores. In this work, we mainly focus on terabyte-scale graphs, where the GPU memory cannot store such a large proportion of features in most cases.

%% file: context/relatedwork.tex
\section{Related Work} 
To our knowledge, \SystemName{} is the first out-of-core GNN training system that trains GNN on terabyte-scale graphs with in-memory efficiency. In the following, we contrast \SystemName{} and existing works in the following aspects. 

% \noindent{\textbf{GNN Frameworks. }
% Several GNN systems~\cite{jia2020improving, wang2019deep,fey2019fast, zhu2019aligraph, lin2020pagraph, yang2022gnnlab, torchquiver, ma2019neugraph,  wang2021gnnadvisor, gandhi2021p3, thorpe2021dorylus, zheng2022distributed, sun2023legion, zhang2023boosting} have emerged in recent years. 
% Most of these GNN systems are built on top of deep learning frameworks like Pytorch~\cite{paszke2019pytorch}, TensorFlow~\cite{abadi2016tensorflow} and MXNet~\cite{chen2015mxnet}. 
% }

\noindent{\textbf{Large-scale GNN Systems.} 
To train large-scale graphs, there are many GNN systems that have been proposed including distributed systems~\cite{liu2021bgl, zheng2020distdgl, zheng2021distributed, gandhi2021p3, zhu2019aligraph, thorpe2021dorylus, zhang2023boosting, zheng2022bytegnn, wang2022neutronstar, zhang2020agl, li2022hyperscale, peng2022sancus, wan2023scalable, demirci2022scalable} and disk-based systems~\cite{waleffe2023mariusgnn, park2022ginex, park2023accelerating}.
Distributed systems leverage multiple machines' host memories to store large-scale graphs while disk-based systems utilize disks as graph storage. 
Ginex~\cite{park2022ginex} and MariusGNN~\cite{waleffe2023mariusgnn} are two disk-based systems that use CPU to manage the disk IO but have CPU bottleneck and low GPU utilization due to issues I$_1$ and I$_2$. GIDS~\cite{park2023accelerating} is a GPU-managed disk-based system but has low training throughput due to issues I$_3$, I$_4$, and I$_5$. 
In contrast, \SystemName{} fully overlaps training and data access while leveraging efficient SSD control and cache design, achieving high throughput similar to in-memory systems. 
}

\noindent\textbf{GPU-initiated Direct SSD Access. } NVMMU~\cite{nvmmu} and Gullfoss~\cite{gullfoss} enable direct GPU-SSD data transfer using the GPUDirect~\cite{gpudirect} technology. Based on these works, SPIN~\cite{spin} and GRAGON~\cite{dragon} introduce in-memory page caches to enhance the data access efficiency and transparency, while GAIA~\cite{gaia} and GPM~\cite{gpm} further move the cache to GPU memory. HippogriffDB~\cite{hippogriffdb} utilizes direct GPU-SSD data transfer to optimize database queries. Morpheus~\cite{morpheus} and GPUKV~\cite{gpukv} optimize data serialization and KV store applications using in-storage computation approaches. All the aforementioned works adopt CPU-initiated SSD accesses. BaM~\cite{qureshi2023gpu} proposes GPU-initiated SSD access in a synchronous manner. In contrast, \SystemName{} utilizes a lightweight asynchronous GPU-initiated IO stack to achieve high disk IO throughput efficiency for GNN training. 

\noindent\textbf{Cache-enabled GNN Systems. } 
Many GNN systems~\cite{lin2020pagraph, torchquiver, yang2022gnnlab, min2022graph, zhang2023ducati, sun2023legion, liu2021bgl, park2023accelerating, park2022ginex} have explored cache design to accelerate GNN training. PaGraph~\cite{lin2020pagraph}, Quiver~\cite{torchquiver}, GNNLab~\cite{yang2022gnnlab}, Data-tiering~\cite{min2022graph}, DUCATI~\cite{zhang2023ducati} and Legion~\cite{sun2023legion} focus on caching the hottest data in GPU memory to minimize data transfer from CPU memory to GPU memory. 
GIDS~\cite{park2023accelerating} also leverages GPU cache but aims to minimize the disk IO.
Ginex~\cite{park2022ginex} regards CPU memory as a cache to mitigate the disk IO.
BGL~\cite{liu2021bgl} explores CPU memory and GPU memory as cache to minimize the network and PCIe traffic and manages the cache by CPU. 
In contrast, \SystemName{} extends the cache hierarchy to heterogeneous CPU and GPU memory and manages cache by GPU to minimize the management overhead.

%% file: context/conclusion.tex
\section{Conclusion}
In this work, we propose \SystemName{}, a system that can train GNN on terabyte graphs in a single machine while achieving throughput comparable with storing all graph data in memory. To achieve this, we first introduce a GPU-initiated asynchronous disk IO stack, allowing the GPU to directly access to graph data on disk. This design requires about 30\% GPU cores to reach the almost maximal disk IO throughput and wastes no GPU cores between IO submission and IO completion to leave the majority of GPU cores for other GNN kernels. Second, we design a GPU-managed heterogeneous cache that extends the cache hierarchy to heterogeneous CPU and GPU memory, mitigating the low throughput of disk IO and enhancing cache lookup throughput significantly. Finally, we build a deep GNN-aware pipeline that seamlessly integrates the computation and communication phases of the entire GNN training process, maximizing the utility of GPU computation cycles. 
Experimental results demonstrate that \SystemName{} can match the training throughput of in-memory GNN systems, even for terabyte-scale graphs. Remarkably, \SystemName{} surpasses the state-of-the-art GPU-managed baselines by up to 6.43$\times$ and exceeds CPU-managed baselines by over a staggering 182$\times$.